\documentclass{article}
\pdfoutput=1
\usepackage{jheppub}

\usepackage{afterpage}

\usepackage{amssymb}
\usepackage{amsmath}
\usepackage{graphicx}
\usepackage{longtable}
\usepackage{verbatim}
\usepackage{amsfonts}
\usepackage{color, colortbl}
\definecolor{LightCyan}{rgb}{0.88,1,1}
\definecolor{LightViolet}{rgb}{1, 0.88,1}
\definecolor{LightGreen}{rgb}{0.88, 1, 0.88}

\usepackage{color, colortbl}
\definecolor{LightCyan}{rgb}{0.88,1,1}
\definecolor{LightViolet}{rgb}{1, 0.88,1}
\definecolor{LightGreen}{rgb}{0.88, 1, 0.88}

\newcommand{\bear}{\begin{array}}
\newcommand{\ear}{\end{array}}

\newcommand{\beq}{\begin{eqnarray}}
\newcommand{\eeq}{\end{eqnarray}}
\newcommand{\beqa}{\begin{eqnarray}}
\newcommand{\eeqa}{\end{eqnarray}}

\def\OMIT#1{{}}
\newcommand{\lsim}{\mathrel{\rlap{\lower4pt\hbox{\hskip1pt$\sim$}}
    \raise1pt\hbox{$<$}}}         
\newcommand{\gsim}{\mathrel{\rlap{\lower4pt\hbox{\hskip1pt$\sim$}}
    \raise1pt\hbox{$>$}}}         

\newcommand{\be}{\begin{equation}}
\newcommand{\ee}{\end{equation}}
\newcommand{\ba}{\begin{eqnarray}}
\newcommand{\ea}{\end{eqnarray}}

\def\lsim{\mathrel{\rlap{\lower4pt\hbox{\hskip1pt$\sim$}}
    \raise1pt\hbox{$<$}}}         
\def\gsim{\mathrel{\rlap{\lower4pt\hbox{\hskip1pt$\sim$}}
    \raise1pt\hbox{$>$}}}         

\begin{document}

\vspace*{-30mm}

\title{\boldmath Probing Charged Matter Through $h \to\gamma\gamma$, Gamma Ray Lines, and EDMs}

\author{JiJi Fan and Matthew Reece}
\affiliation{Department of Physics, Harvard University, Cambridge, MA 08540, USA}

\vspace*{1cm}

\abstract{Numerous experiments currently underway offer the potential to indirectly probe new charged particles with masses at the weak scale. For example, the tentative excess in $h \to \gamma \gamma$ decays and the tentative gamma-ray line in Fermi-LAT data have recently attracted attention as possible one-loop signatures of new charged particles. We explore the interplay between such signals, dark matter direct detection through Higgs exchange, and measurements of the electron EDM, by studying the size of these effects in several models. We compute one-loop effects to explore the relationship among couplings probed by different experiments. In particular, models in which dark matter and the Higgs both interact with charged particles at a detectable level typically induce, at loop level, couplings between dark matter and the Higgs that are around the level of current direct detection sensitivity. Intriguingly, one-loop $h \to \gamma \gamma$ and ${\rm DM}~{\rm DM} \to \gamma \gamma$, two-loop EDMs, and loop-induced direct detection rates are all coming within range of existing experiments for approximately the same range of charged particle masses, offering the prospect of an exciting coincidence of signals at collider, astrophysical, underground and atomic physics measurements.}
\maketitle

\section{Introduction}
\begin{figure}[!h]\begin{center}
\includegraphics[width=0.8\textwidth]{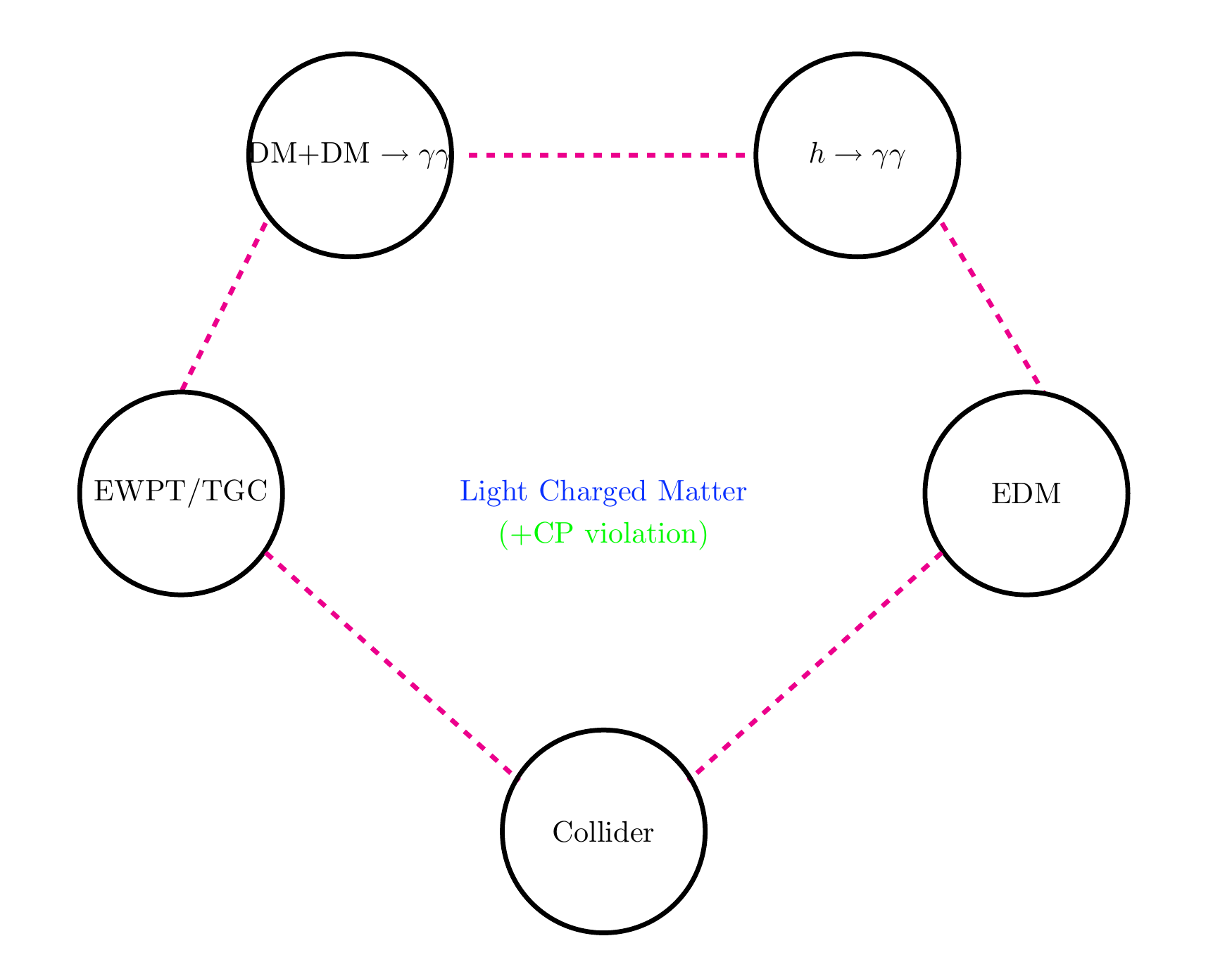}
\end{center}
\caption{Possible experimental signals of weak-scale colorless charged matter. }
\label{fig:setup}
\end{figure}

It has been expected for a long while that new physics would proliferate at about the electroweak scale. While the discovery of a Higgs-like particle at the LHC~\cite{CMS,ATLAS} is a profound advance in particle physics, there is so far no evidence for other new particles beyond the Standard Model (SM). Although new colored particles are strongly constrained, color singlet new particles could still be at large, as the LHC has just begun to be sensitive to electroweak processes. It is interesting to explore all possible experimental consequences of new light colorless weakly-interacting charged particles. Specifically, depending on their couplings, they could possibly lead to the following observations, summarized in Fig.~\ref{fig:setup}:
\begin{itemize}
\item{Features in the photon spectrum from the galactic center or other astrophysical sources if the charged matter couples to the dark matter (DM).}
\item{Modification of Higgs decay, in particular, $h \to \gamma\gamma$, if they couple to the Higgs.}
\item{Electron or neutron EDM if there is CP violation in their couplings.}
\item{Modifications of electroweak precision observables if they are weakly charged. Certain parameters such as oblique parameters are stringently constrained but current bounds on triple gauge couplings (TGCs) are still weak.}
\end{itemize}
Light charged particles could also be produced through electroweak processes at the LHC~\cite{ArkaniHamed:2012kq, Batell:2012mj}, leading to different signals depending on their decay modes. It has been shown by some rough estimates that such particles could be within reach of discovery in almost all cases in the 8 TeV run at the LHC, and in even the most difficult cases at 14 TeV~\cite{ArkaniHamed:2012kq}. However, in this paper, we will not discuss the direct detection of these charged particles at colliders. The purpose of this paper is to explore the constraints on different indirect observables listed above and possible correlations between them in scenarios with light charged matter. In particular, we are interested in two correlations in the cases: a) if the charged matter is coupled to both DM and Higgs; b) if there is an order one phase in the charged matter sector that cannot be rotated away.

Now we would like to review briefly all the possible observations and their current experimental status. If the charged matter couples to DM, it could mediate loop-level annihilations of DM particles into two photons (and possibly photon + $Z$). This possibility becomes interesting as recently an observation has been made of gamma-ray line emissions in the galactic center using 3.7 years of Fermi-LAT data~\cite{Bringmann:2012vr1, Weniger:2012tx}. While the feature is consistent with one single line at about 128 GeV, a pair of two lines at around 111 GeV and 128 GeV gives a slightly better fit to data~\cite{Rajaraman:2012db, Su:2012ft}. (It is unclear if this continues to be true in Fermi's updated Pass 8 data, of which only limited information is publicly available at this time. It is known that the higher-energy line shifted to 135 GeV after recalibration, and unclear whether the second line can still be detected. It is our understanding that many details may still change as the Pass 8 data is validated and analyzed before its public release, so we will take a wait-and-see attitude.) The Harvard group further reported possible double line emissions at about 111 and 128 GeV from unassociated Fermi-LAT sources~\cite{Su:2012zg}. These lines could be consistent with DM particles annihilating at one loop to $\gamma \gamma$ (and $\gamma Z$ for two lines) with considerable cross sections of order $\langle \sigma v \rangle \sim 10^{-27}$ cm$^{3}$s$^{-1}$~\cite{Weniger:2012tx, Tempel:2012ey}. On the other hand, a similar signal in ``Earth limb photons'' is difficult to reconcile with a dark matter interpretation. In any case, the current cross section at which gamma ray lines are potentially detectable is $10^{-27}$ cm$^3$s$^{-1}$, and whether the tentative lines persist or not, it is interesting to consider interpretations of hypothetical new physics at the current sensitivity threshold.

It is non-trivial to have a DM model give rise to these lines without being constrained by other observations. For example, MSSM neutralino DMs annihilating into two photons are ruled out by the continuous gamma ray spectrum as at tree-level level, they annihilate into other SM particles such as $WW$ and $ZZ$, whose subsequent decays could also produce photons with continuous energy. More specifically, fixing the loop level cross section to diphotons to fit the signal, cross sections of any processes contributing to the photon continuum spectrum are constrained to be no more than 5 to 10 times the loop level cross section~\cite{Buchmuller:2012rc, Cohen:2012me, Cholis:2012fb}. 
Given that, the simplest possibility is that DM is a SM gauge singlet annihilating through a loop of light SM charged particles. To get the desired rate, this requires quite large couplings and numerical enhancements from coincidences in the mass of DM and the particle running in the loop~\cite{Cline:2012nw, Buckley:2012ws, Weiner:2012cb,Bergstrom:2012bd,Wang:2012ts,Weiner:2012gm, Baek:2012ub,D'Eramo:2012rr, Farzan:2012kk, Choi:2012ap, Kopp:2013mi}. Another possible topology for a rate enhancement is through an $s$-channel exchange of a pseudo-scalar or vector~\cite{Buckley:2012ws, Lee:2012bq, Das:2012ys, Tulin:2012uq, Dudas:2012pb,Lee:2012wz,Chalons:2012xf,SchmidtHoberg:2012ip,Bai:2012yq, Bai:2012qy, Kopp:2013mi} with mass tuned close to twice the DM mass. The pseudo-scalar (or vector) couples through a loop of light electrically charged particles to two photons (or a photon and $Z$).\footnote{Two notably different options exist in which the gamma ray lines are actually different shapes with narrow widths smaller than the Fermi-LAT current resolution. One is that the lines are narrow box-shaped features from a process like ${\rm DM} + {\rm DM} \to \pi_0^h + \pi_0^h$ in which $\pi_0^h$ subsequently decay to $\gamma\gamma$ and $\gamma Z$~\cite{Fortin:2009rq, Ibarra:2012dw, Chu:2012qy}. The narrow width of the box could be explained if $\pi_0^h$ is degenerate with DM in mass. A simple elegant explanation could be that the $\pi_0^h$ are in the same multiplet with DM due to a symmetry~\cite{Fan:2012gr}. A UV completion of this scenario was proposed in Ref.~\cite{Cvetic:2012kj}. The other option is internal bremsstrahlung, in which DM annihilating with a $t$-channel particle emits photons with energy below the DM mass~\cite{Bringmann:2007nk}. When the mass of the $t$-channel particle is tuned to be close to the DM mass, the edge could be peaked around DM mass~\cite{Bringmann:2012vr, Shakya:2012fj}. }

If the colorless charged matter couples to the Higgs, it will modify the $h\gamma\gamma$ coupling at the one-loop level. Originally, both CMS and ATLAS observed an enhancement in the Higgs decaying to diphoton rate while the Higgs decaying to diboson rates $\sigma \times Br(h\to WW^*, ZZ^*)$ have been roughly consistent with the SM Higgs expectation~\cite{CMS, ATLAS}. The most recent update is that the most sensitive CMS analysis observes a $\sigma \times Br(h\to\gamma\gamma)$ of $0.78 \pm 0.27$ times the SM rate~\cite{CMSdiphotonupdate}, while ATLAS observes $1.65 \pm 0.24{(\rm stat)}^{+0.25}_{-0.18}{\rm(syst)}$ times the SM rate~\cite{ATLASdiphotonupdate}. Given these numbers, it is entirely possible either that the Standard Model value will prove to be correct, or that a moderate enhancement would persist. Enhancements at the level suggested by the ATLAS central value would require new charged matter with mass close to the LEP bound, 100 GeV~\cite{Carena:2012xa, Joglekar:2012vc, ArkaniHamed:2012kq}. It should be emphasized that these new charged particles should not be colored; otherwise they will be ruled out by vacuum instability constraints~\cite{ArkaniHamed:2012kq, Reece:2012gi}. For further recent studies on possible mechanisms for a diphoton enhancement without enhancing diboson rate, see also Refs.~\cite{Carena:2011aa,Carena:2012gp,Almeida:2012bq,Davoudiasl:2012ig,Bae:2012ir,Kitahara:2012pb,Kobakhidze:2012wb,Chun:2012jw,Gogoladze:2012jp,Moreau:2012da,Chala:2012af,Choi:2012he,Batell:2012zw,Davoudiasl:2012tu,Carena:2012mw,Huo:2012tw,Huang:2012rh,Berg:2012cg,An:2012vp}. A good, up-to-date review of the status of Higgs physics may be found in Ref.~\cite{Azatov:2012qz}.

On the other hand, if the diphoton enhancement disappears, the data would constrain the coupling of a light charged particle to the Higgs, or more accurately, the dependence of its mass on electroweak symmetry breaking. 

Furthermore, generically light charged matter coupling to the Higgs can have an order one CP violating phase~\cite{Voloshin:2012tv}, leading at two loops to a signal already constrained by electron electric dipole moment (EDM) experiments~\cite{McKeen:2012av}. In the near future, there will be an update of the electron EDM experiment which could potentially improve the current bound by one order of magnitude~\cite{Vutha:2009ux}. Thus EDMs could be an independent restriction on possible deviations of the $h\gamma\gamma$ coupling. 

Depending on the quantum numbers, the light charged matter could contribute to both the oblique parameters in the electroweak precision analysis and non-oblique parameters such as triple gauge couplings (TGC). For the constraints on the oblique parameters, it has been shown in~\cite{Joglekar:2012vc, ArkaniHamed:2012kq} that they could be evaded easily. In this paper, we will elaborate more on the constraints on TGCs. 

One topic we will not discuss is muon $g-2$. Although formally it fits very well with our theme, since magnetic dipole moments are essentially the CP-conserving partners of electric dipole moments, phenomenologically its status is somewhat different. The difficulty is that the measured and theoretical values disagree, $\delta a_\mu \approx (2.8 \pm 0.8) \times 10^{-9}$~\cite{Bennett:2006fi,Hagiwara:2011af,Davier:2010nc}, and this discrepancy is too large to correspond to a two-loop effect analogous to the EDMs we discuss, which translate to $\delta a_\mu$ on the order of $10^{-12}$. Hence, any two-loop effect is either masked by other new physics that explains the discrepancy (a very exciting possibility), or, more plausibly, hidden beneath theoretical and experimental uncertainties that remain to be resolved. A one-loop new physics contribution could explain the data, but this requires physics like sleptons that carry lepton flavor quantum numbers; for instance, a recent discussion relating $g-2$ to $h \to \gamma\gamma$ via sleptons appeared in Ref.~\cite{Giudice:2012pf}. We will have nothing to add to its discussion, because throughout this paper, we discuss only new charged particles that carry no flavor quantum numbers.

In Section~\ref{sec:HiggsDM}, we will discuss two models that introduce new charged particles running in loops to fit a large $h\to \gamma\gamma$ excess and dark matter annihilation to a gamma-ray line. In both cases, we show that loops can induce couplings of the Higgs to dark matter that are in tension with direct detection limits. (A third such model is discussed in Appendix~\ref{app:boxdiagram}, to avoid tedious repetition in the main text.) In Section~\ref{sec:CPodd}, we discuss two-loop EDM effects and the expected correlation between new physics in CP-even and CP-odd observables if phases are generic. (A minor technical detail is discussed in Appendix~\ref{app:CPoddops}.) Sections~\ref{sec:HiggsDM} and~\ref{sec:CPodd} have some overlap in their content, particularly in discussions of corrections to $h \to \gamma\gamma$ from new vectorlike fermions, but are written so that they can be read independently. Concluding remarks are offered in Section~\ref{sec:conclusions}.

\section{Induced Higgs-DM coupling via charged matter}
\label{sec:HiggsDM}
\begin{figure}[!h]\begin{center}
\includegraphics[width=0.4\textwidth]{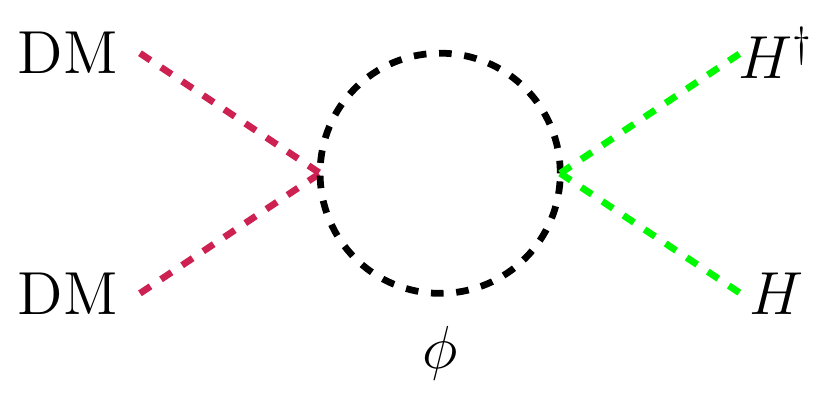} \quad
\includegraphics[width=0.45\textwidth]{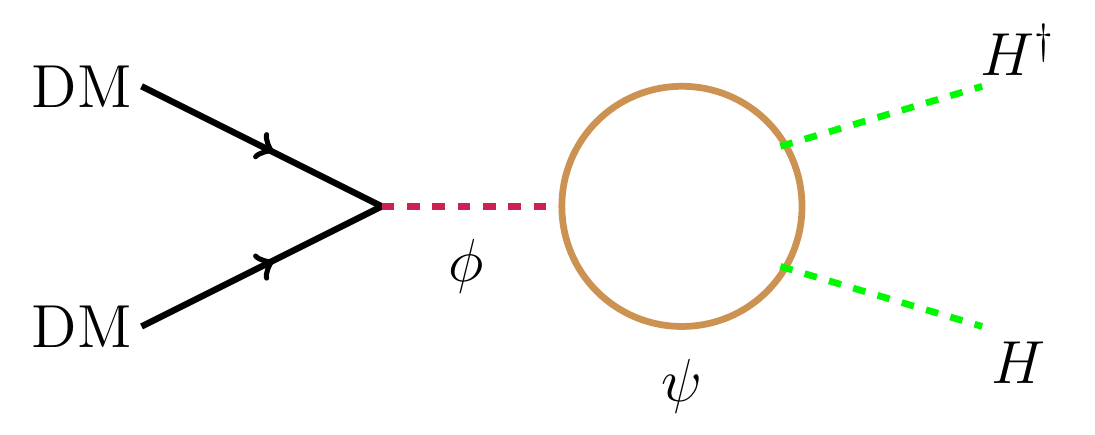} \\
\includegraphics[width=0.4\textwidth]{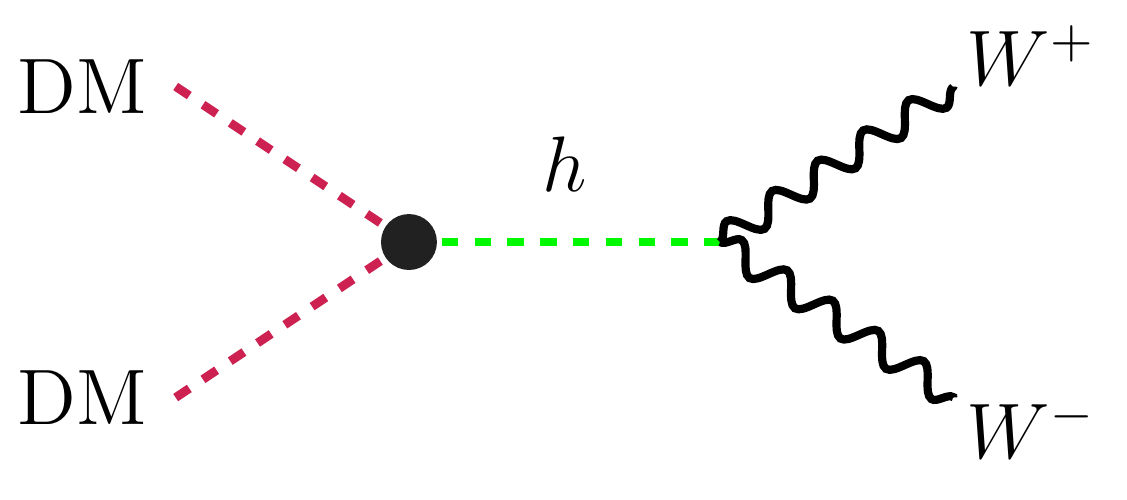} \quad
\includegraphics[width=0.4\textwidth]{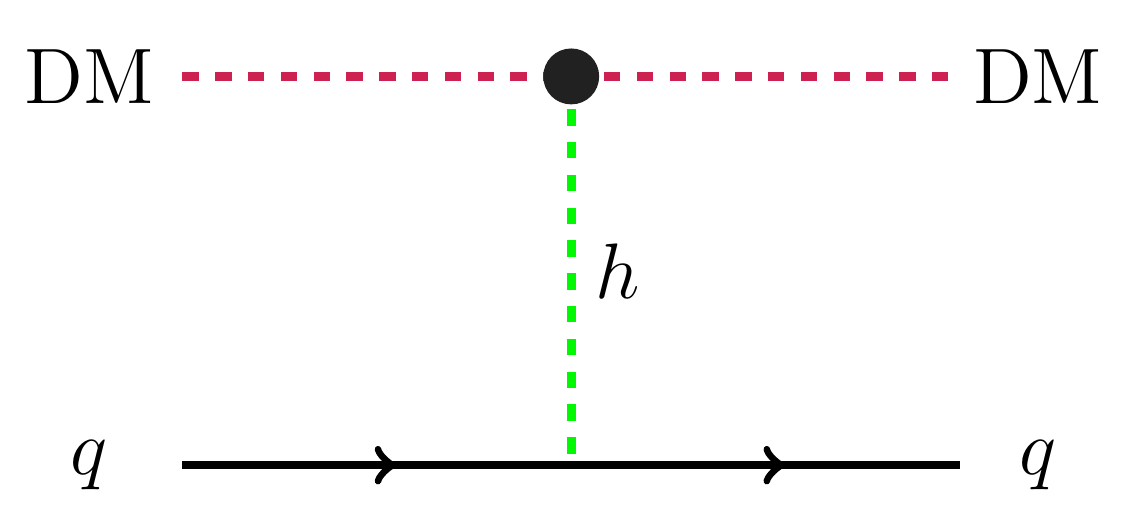} 
\end{center}
\caption{Upper: Possible diagrams contributing to an induced coupling between DM and Higgs. Lower: Processes leading to photon continuum and scattering with nucleons at direct detection. The blobs represent (one-loop) induced coupling between DM and Higgs through charged matter.}
\label{fig:constraints}
\end{figure}

Because the Fermi-LAT gamma ray line and the $h \to \gamma \gamma$ rate could both be taken as hints of new charged matter at the weak scale, it is tempting to postulate new particles that explain both possible signals. Indeed, such a suggestion has been made in Refs.~\cite{Cline:2012nw,Baek:2012ub,SchmidtHoberg:2012ip}. However, when considering the full effective theory of the Higgs, dark matter, and new charged particles, one must be careful: couplings between dark matter and the Higgs are constrained by direct detection experiments (${\rm DM} + q \to {\rm DM} + q$) as well as (depending on the CP properties of the initial state) indirect detection DM + DM $\to h^* \to WW, ZZ$. Even if we initially assume that dark matter and the Higgs are not directly coupled, renormalization group evolution in the effective theory will inevitably generate couplings between them. The relationship among these various processes is illustrated in Fig.~\ref{fig:constraints}.

In considering dark matter models that can achieve the Fermi-LAT gamma ray line rate, we will not be careful to select points in parameter space that can achieve a thermal relic abundance. We refer the reader to Ref.~\cite{Tulin:2012uq} for a discussion of the thermal history in the models we consider. These models are, for appropriate choices of masses of the additional particles, capable of fitting the Fermi-LAT line rate and having a thermal relic abundance, and our comments apply to those parameter choices. But, given our ignorance of the thermal history of the universe before BBN, and the plausibility of some nonthermal dark matter scenarios, we choose to consider the parameter space more broadly and not single out points that have a thermal relic abundance.

\subsection{All-scalar effective theory}
\label{sec:allscalar}

\begin{figure}[!h]\begin{center}
\includegraphics[width=0.4\textwidth]{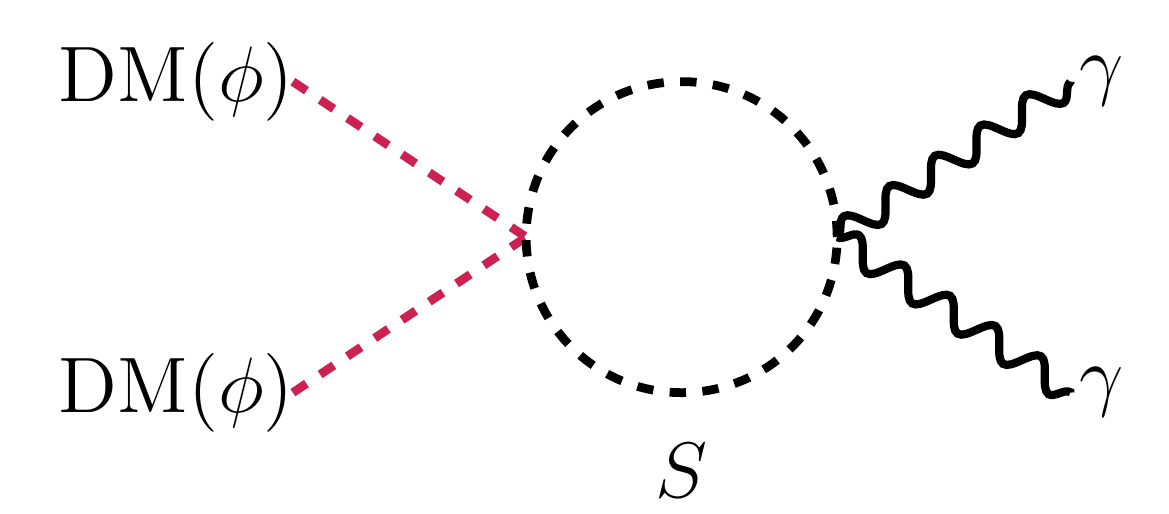} 
\quad \includegraphics[width=0.4\textwidth]{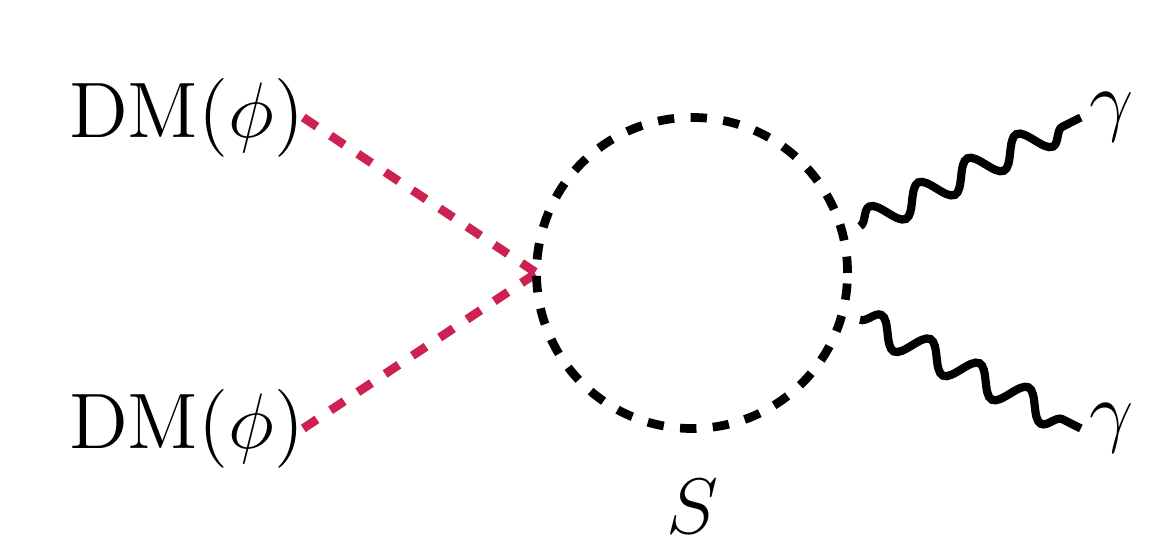} 
\end{center}
\caption{The annihilation processes leading to 2 photons.}
\label{fig:process1}
\end{figure}

First, we will illustrate the idea in the context of a simple model in which the dark matter and the new charged particles are all scalars. This is essentially the model studied in Ref.~\cite{Cline:2012nw,Baek:2012ub}, although some of our conclusions differ. (The DM/Higgs interactions were discussed much earlier in Ref.~\cite{Burgess:2000yq}.) We take DM to be a real SM singlet scalar $\phi$, with $m_\phi = 130$ GeV and a ${\mathbb Z}/2$ symmetry $\phi \to -\phi$. The charged matter is a scalar $S$ charged under the SM gauge group as $(1,N)_Y$ with $N_S$ species and a common mass $m_S$. The relevant interactions are 
\beq
-{\cal L} & \supset & \lambda_{\phi S} \phi^2 |S|^2 + \lambda_{HS} |S|^2 |H|^2 + \lambda_{\phi H} \phi^2 |H|^2 + m_{S;0}^2 |S|^2 + \lambda_S |S|^4 + \frac{1}{2}m_\phi^2 \phi^2 + \lambda_\phi \phi^4 \nonumber \\
& - &  \mu_H^2 |H|^2 + \lambda_H |H|^4
\eeq
with the annihilation processes into two photons depicted in Fig.~\ref{fig:process1}. In the case of more than one $S$ species, $|S|^2$ should be interpreted as $\sum_i |S_i|^2$ and $|S|^4$ as $\left(\sum_i |S_i|^2\right)^2$; one could consider more general contractions of the $S$ flavor indices, but there would be no qualitatively different physics. Here $\mu_H = \frac{1}{\sqrt{2}} m_h$ is fixed by the measured Higgs mass $m_h \approx 125$ GeV, which together with the measured Higgs VEV also determines $\lambda_H \approx 0.13$. The physical $S$ mass is given by
\beq
m_S^2 = m_{S;0}^2 + \frac{1}{2} \lambda_{HS}v^2.
\eeq
(In the case that $S$ carries $SU(2)_W$ quantum numbers, additional couplings may be present, e.g. $(H^\dagger S)(S^\dagger H)$ where $SU(2)_W$ indices are contracted within the parentheses. We will not discuss the full parameter space of such couplings, which we expect would not qualitatively change any of our conclusions.) Thanks to the Higgs low-energy theorem~\cite{Ellis:1975ap, Shifman:1979eb}, we see that we require $\lambda_{HS} < 0$ if loops of the $S$ field are to increase the $h \to \gamma \gamma$ rate. In order to prevent the potential from being unbounded from below due to this negative quartic, we require
\beq
\lambda_S \geq \lambda_{S;min} \equiv \frac{\lambda_{HS}^2}{4\lambda_H};
\eeq 
allowing a metastable, rather than absolutely stable, vacuum ameliorates this constraint by about a factor of 2, according to a tree-level calculation of the bounce action for vacuum decay~\cite{Reece:2012gi}.

\subsubsection{The constraint from the gamma ray continuum}
\label{subsec:continuum}
The interaction $\lambda_{\phi H} \phi^2 |H|^2$ provides a dark matter annihilation channel DM + DM $\to h^* \to WW, ZZ$ with cross section given by~\cite{Burgess:2000yq} 
\beq
\langle \sigma v \rangle& = &\sum_{i=W,Z} n_i \frac{\left|\lambda_{\phi H}\right|^2}{2 \pi m_\phi^2}\sqrt{1-\frac{m_i^2}{m_\phi^2}}\frac{m_i^4}{\left(4m_\phi^2-m_h^2\right)^2}\left(2+\frac{(2m_\phi^2-m_i^2)^2}{m_i^4}\right) \\
&=&\left|\frac{\lambda_{\phi H}}{0.028} \right|^2 3\times 10^{-26} {\rm cm}^3{\rm s}^{-1},
\eeq
taking $m_\phi = 130$ GeV and $m_h = 125$ GeV. In the first line $n_i = 1$ for $W$ bosons and $1/2$ for $Z$ bosons is the Bose factor in the case of identical final state particles. There is also a phase-space suppressed annihilation to the $hh$ final state. Note that $\lambda_{\phi H}$ in our notation corresponds to what was denoted $\lambda_{hX}/2$ in Ref.~\cite{Cline:2012nw}. Based on studies of continuum gamma rays from the galactic center in Refs.~~\cite{Buchmuller:2012rc, Cohen:2012me, Cholis:2012fb,Hooper:2012sr} (as well as comparably strong constraints from radio in Ref.~\cite{Laha:2012fg}), it appears safe to say that an annihilation rate of $10^{-25}~{\rm cm}^3{\rm s}^{-1}$ to $WW$ and $ZZ$ is ruled out even with conservative assumptions about astrophysical backgrounds, while a slightly more aggressive approach to the data would extend the limit down to around 1 to $2 \times 10^{-26}~{\rm cm}^3{\rm s}^{-1}$. We will quote the bound as:
\beq
\left|\lambda_{\phi H}\right| \lsim 0.05.
\eeq
Note that different models could shut off this indirect detection channel; for example, Majorana fermion DM is in a CP-odd initial state when annihilating, so annihilation through an off-shell CP-even Higgs is suppressed.

\subsubsection{Direct detection constraint}
\label{subsec:direct}
The cross section of the scalar DM $\phi$ scattering off a nucleon through Higgs exchange is 
\beq
\sigma_{SI}&=&\frac{\left|\lambda_{\phi H}\right|^2 m_n^4 f^2}{\pi m_h^4 m_\phi^2} \\
&=& \left(\frac{\lambda_{\phi H}}{0.05}\right)^2 5\times 10^{-45} {\rm cm}^2,
\eeq
where we take the nucleon mass $m_n = 0.94$ GeV and $f$ parametrizes the nucleon matrix element
\beq
\langle n |m_q \bar{q}q| n\rangle\equiv f_q m_n [\bar{n}n], \quad f= \sum_{q=u,d,s,c,b,t}f_q=\frac{2}{9}+\frac{5}{9}\sum_{q=u,d,s}f_q.
\eeq
We use $f=0.30$ which is the central value obtained from the analysis in~\cite{Giedt:2009mr}. The most recent update of Xenon 100 constrains $\sigma_{SI}$ to be smaller than $3\times 10^{-45} {\rm cm}^2$ for DM mass at 130 GeV~\cite{Aprile:2012nq}. A similar constraint can be obtained for fermionic DM scattering off the nucleon through Higgs exchange.

In summary, both photon continuum and direct detection constrain the induced DM Higgs coupling to be smaller (in absolute value) than about 0.05. 

\subsubsection{Matching the data}

\begin{figure}[!h]\begin{center}
\includegraphics[width=\textwidth]{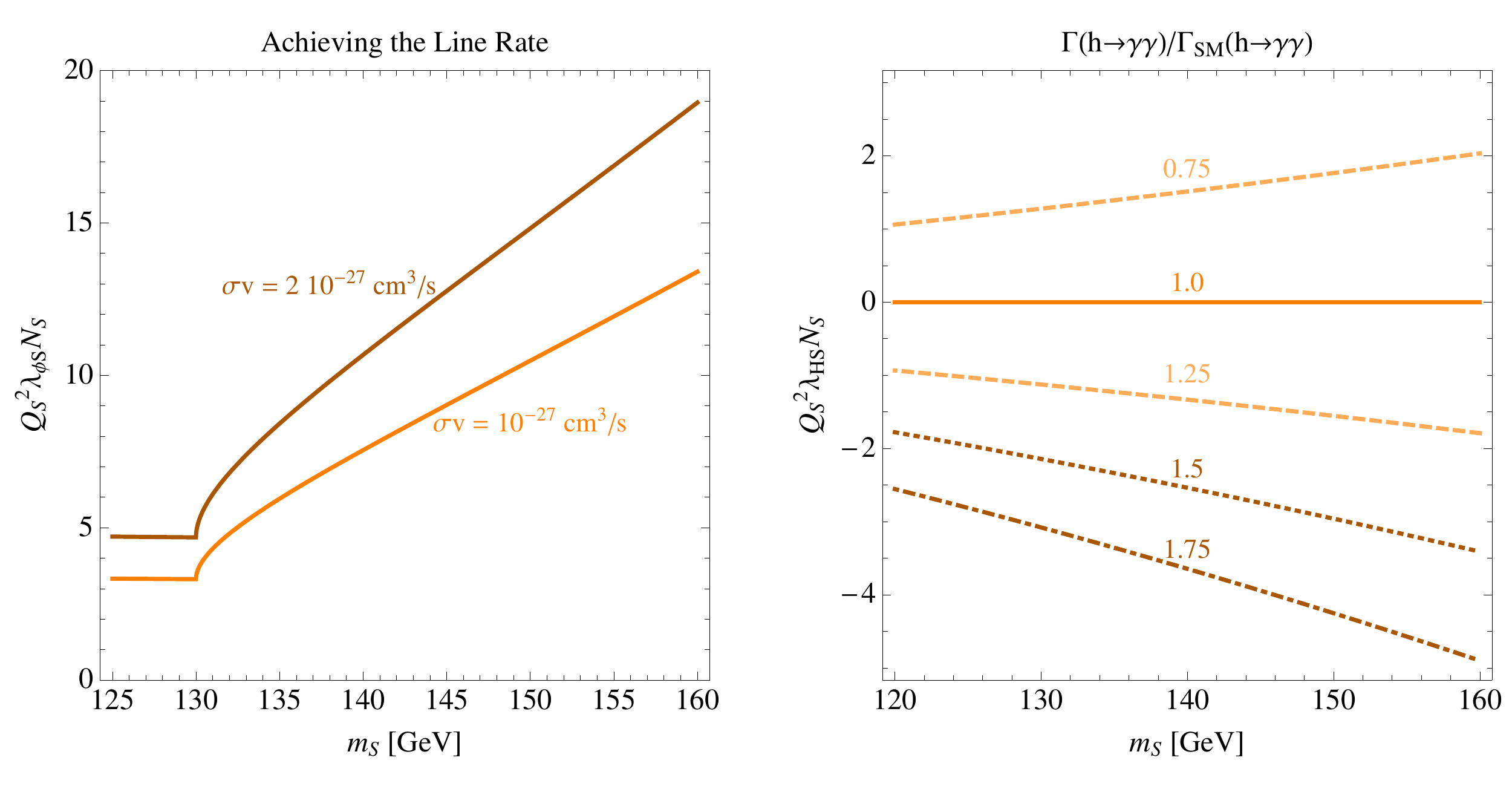}
\end{center}
\caption{Left: Contours of $(\sigma( \phi \phi \to \gamma \gamma) v) = 1,2 \times 10^{-27}$ cm$^{3}$s$^{-1}$ in the ($m_S, \lambda_{\phi S}N_S\sum Q_s^2$) plane. Right: Contours of $\mu_{\gamma\gamma}$ in the ($m_S, \lambda_{H S} N_S\sum Q_s^2$) plane. }
\label{fig:scalar}
\end{figure}

The cross section of the DM annihilation to diphotons is
\beq
(\sigma v)( \phi \phi \to \gamma \gamma)  = \frac{1}{32 \pi^3 m_\phi^2} \left|\alpha \lambda_{\phi S} N_S \left(\sum Q_s^2\right)  \tau_\phi^{-1} A_0(\tau_\phi)\right|^2,
\eeq
where $\sum Q_s^2$ sums the charge squared over all components inside $S$ and 
\beq
A_0(\tau)= - \tau + \tau^2 f(\tau^{-1}) \quad {\rm with}\; \tau_\phi= m_S^2/m_\phi^2, \;\; f(x)={\rm arcsin}^2\sqrt{x}.
\eeq
(For $m_S < m_\phi$, it is necessary to analytically continue $f(x)$; see, for instance, the discussion of $h \to \gamma\gamma$ in Ref.~\cite{Gunion:1989we}.) Demanding $(\sigma v)( \phi \phi \to \gamma \gamma) = 10^{-27}$ cm$^{3}$s$^{-1}$, one obtains $\lambda_{\phi S} N_S \sum Q_s^2$ as a function of $\delta_m\equiv m_S-m_\phi$. The result is presented in the left panel of Fig.~\ref{fig:scalar}. 

At one-loop level, DM could also annihilate into $\gamma Z, ZZ$ final states with the first final state leading to a second line at $m_\phi-m_Z^2/(4m_\phi) = 114$ GeV. If the charged matter transforms non-trivially under $SU(2)_W$, DM also annihilates into $WW$, which together with the $ZZ$ final state, contributes to the photon continuum. Too large a continuum rate relative to photons would be excluded by data. It is straightforward to calculate results for general quantum numbers, but we will simply quote the case where the charged scalar $S$ has only hypercharge. The ratio of other annihilation signals to the $\gamma\gamma$ line is:
\beq
\frac{\sigma(Z\gamma)}{\sigma(\gamma\gamma)}&=&2\tan^2\theta_W\left(1-\frac{m_Z^2}{4m_\phi^2}\right)^3\approx 0.4\\
\frac{\sigma(ZZ)}{\sigma(\gamma\gamma)}&=&2\tan^4\theta_W\left(1-\frac{m_Z^2}{m_\phi^2}\right)^{3/2}\approx 0.006.
\eeq
These formulas can have significant corrections from loop functions if the charged scalar is lighter than 130 GeV.

Finally, the charged scalars contribute to the $h\to \gamma\gamma$ rate
\beq
\mu_{\gamma\gamma}=\frac{\sigma\times Br(h\to \gamma\gamma)}{{\rm SM}}=\left|1-\frac{\lambda_{HS} N_S\sum Q_s^2}{2}\frac{v^2}{m_S^2}\frac{A_0(\tau_s)}{6.49}\right|^2,
\eeq
where $\tau_s=4m_S^2/m_h^2$ and $- 6.49$ is the SM $h\gamma\gamma$ amplitude. We plot $\lambda_{HS} N_S \sum Q_s^2$ as a function of $m_S$ for several choices of $\mu_{\gamma\gamma}$ ranging from $0.75$ to $1.75$ in the right panel of Fig.~\ref{fig:scalar}. 

There are two things to note about Fig.~\ref{fig:scalar}. The first is that achieving a reasonable fit both to the Fermi-LAT gamma ray line and to an enhanced $h \to \gamma \gamma$ rate requires both $\lambda_{\phi S}$ and $\lambda_{HS}$ to be order-one numbers. The second is that, to fit the gamma ray line, it is necessary that $m_S$ not be much larger than the dark matter mass; otherwise, the coupling needed to achieve a large enough cross section rapidly becomes nonperturbatively large. (This raises the intriguing possibility that the annihilation process $\phi\phi \to S S$, forbidden today if $m_S > m_\phi$, was active in the early universe and played a key role in determining the dark matter relic abundance~\cite{Griest:1990kh,Tulin:2012uq}.) If we fix a small splitting, say $m_S - m_\phi = 1$ GeV, and consider $S$ to be a set of $N_S$ degenerate states of charge 1, then the coupling we need for $\sigma v = 10^{-27}$ cm$^{3}$s$^{-1}$ is already $\lambda_{\phi S} N_S = 4.3$. (Furthermore, avoiding a potential that is unbounded from below requires another large coupling, $\lambda_S \gsim 9.3/N_S^2$.) A 50\% enhancement in the $h \to \gamma\gamma$ signal requires $\lambda_{HS} N_S = -2.2$, and a 25\% enhancement requires $\lambda_{HS} N_S \approx -1.1$. We will now investigate some of the consequences of these rather large couplings.

\subsubsection{RGEs}
\label{subsec:scalarrges}

The one loop RGEs, keeping the scalar quartic couplings, the top Yukawa, and the larger SM gauge coupling effects, are presented below, for the case where $S$ is charged only under hypercharge. (See related recent work in~\cite{Gonderinger:2012rd,Cheung:2012nb}.)
\beq
16\pi^2 \beta(\lambda_H) &=& 24 \lambda_H^2 + 12 \lambda_H y_t^2 + 2 \lambda_{\phi H}^2 + N_S \lambda_{HS}^2 - 6 y_t^4 + \frac{9}{8} g_2^4 - 9 g_2^2 \lambda_H \\
16\pi^2 \beta(\lambda_{\phi H}) &=& 8 \lambda_{\phi H}^2 + 24 \lambda_{\phi H} \lambda_\phi + 12 \lambda_H \lambda_{\phi H} + 2 N_S \lambda_{HS} \lambda_{\phi S} + 6 \lambda_{\phi H} y_t^2 - \frac{9}{2} g_2^2 \lambda_{\phi H} \\
16\pi^2 \beta(\lambda_{HS}) &=& 4 \lambda_{HS}^2 + (4 + 4 N_S) \lambda_{HS} \lambda_S + 4 \lambda_{\phi H} \lambda_{\phi S} + 12 \lambda_H \lambda_{HS} + 6 \lambda_{\phi H} y_t^2 -\frac{9}{2} g_2^2 \lambda_{\phi H} \\
16\pi^2 \beta(\lambda_{\phi S}) &=&  8 \lambda_{\phi S}^2 + (4 + 4 N_S) \lambda_S \lambda_{\phi S} + 4 \lambda_{HS} \lambda_{\phi H} + 24 \lambda_\phi \lambda_{\phi S}\\
16\pi^2 \beta(\lambda_S) &=& (16 + 4 N_S) \lambda_S^2 + 2 \lambda_{HS}^2 + 2 \lambda_{\phi S}^2 \\
16\pi^2 \beta(\lambda_\phi) &=& 72 \lambda_\phi^2 + 2 \lambda_{\phi H}^2 + N_S \lambda_{\phi S}^2\\
16\pi^2 \beta(y_t) &=& \frac{9}{2} y_t^3 - 8 g_3^2 y_t - \frac{9}{4} g_2^2 y_t
\eeq
(An easy way to keep track of the numerical factors appearing the ${\cal O}(\lambda^2)$ terms in beta functions of quartic terms is to notice that they must compensate the $\log\mu$ term in the Coleman-Weinberg potential, so the beta functions amount to reading off coefficients in ${\rm Tr}{\cal M}^4$.) Note, in particular, that we have a simple estimate for a coupling between dark matter and the Higgs induced by a loop of $S$ fields as shown in the upper left panel of Fig.~\ref{fig:constraints}:
\beq
\lambda_{\phi H} \approx \frac{\lambda_{HS} \lambda_{\phi S} N_S}{8\pi^2} \log\frac{\Lambda}{m_S} \approx -0.24 \frac{\lambda_{\phi S} N_S}{4.3} \frac{\lambda_{HS} N_S}{-2.2} \frac{1}{N_S} \frac{\log(\Lambda/m_S)}{2.0}. \label{eq:lambdaphiHestimate}
\eeq
Note that $\log(1~{\rm TeV}/m_S) \approx 2$, so the log will already have this size even when running from quite a low scale. It is apparent that our bound $\left|\lambda_{\phi H}\right| \lsim 0.05$ from direct and indirect detection is in some tension with our desire to explain both the Fermi-LAT gamma ray line and an enhancement in $h \to \gamma\gamma$. Even for only a 10\% enhancement of $\mu_{\gamma\gamma}$, $\lambda_{HS} = -0.5$ (fixing $N_S = 1$) and so $\lambda_{\phi H} \approx -0.05$, still in tension with the bounds in Sec.~\ref{subsec:continuum} and~\ref{subsec:direct}. The problem is ameliorated when the number of species, $N_S$, is large, but this also makes the renormalization group effects large.

For $N_S = 1$, an even more immediate problem is that avoiding a potential unbounded from below requires $\lambda_S \approx 9$. Then the leading term in the $\lambda_S$ beta function is $\frac{1}{16\pi^2} 20 \lambda_S^2 \approx 11$, and there is no sense in which the theory is under perturbative control. For a single charged scalar (of charge 1), it is simply not possible to discuss a 50\% enhancement in $h \to \gamma\gamma$ while maintaining a perturbative theory. Hence, we should focus attention on $N_S \geq 2$. The large couplings suggest that our RGEs will become nonperturbative at low scales. We can quantify this by defining, for each coupling, a perturbativity limit at which the beta function becomes $\geq 1$ when all other couplings are turned off. (This definition was used in Ref.~\cite{Cheung:2012nb}. It usually corresponds to smaller couplings than those for which the two-loop beta function is larger than the one-loop beta function, but in practice we find that a coupling exceeding this bound will very quickly run large enough to exceed the other as well.) For example, we define $\lambda_H^{\rm max}$ by the condition $\frac{1}{16\pi^2} 24 \left(\lambda_H^{\rm max}\right)^2 = 1$, i.e. $\lambda_H^{\rm max} \equiv \sqrt{2/3} \pi \approx 2.6$. For each coupling, we define an analogous $\lambda_i^{\rm max}$ and define a {\em normalized coupling} by ${\bar \lambda}_i = \lambda_i/\lambda_i^{\rm max}$. 

\begin{figure}[!h]\begin{center}
\includegraphics[width=\textwidth]{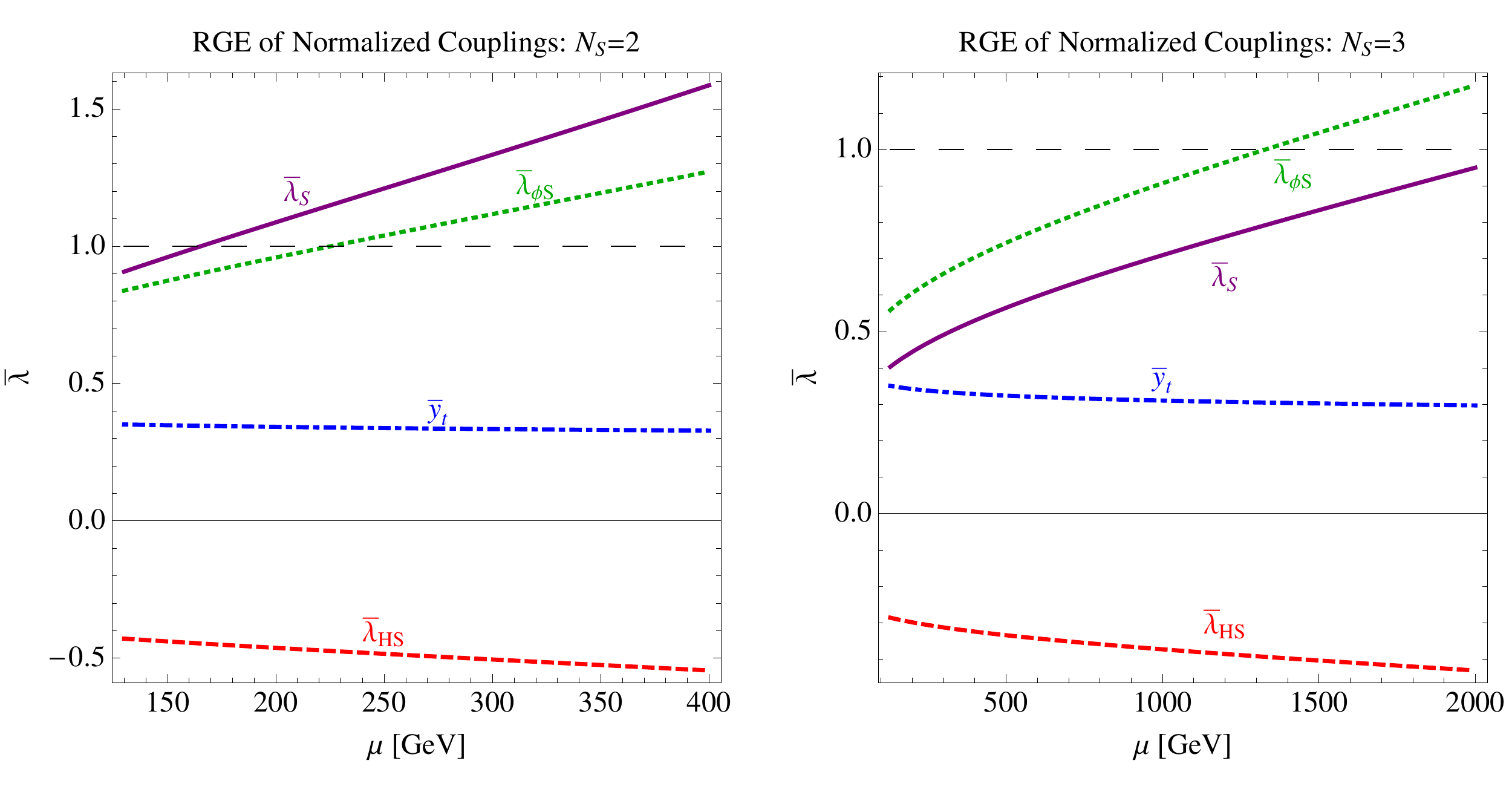}
\end{center}
\caption{Renormalization group evolution of ``normalized couplings,'' with the number of scalar species $N_S = 2$ at left and $3$ at right. The normalized couplings ${\bar\lambda}_i = \lambda_i/\lambda_i^{\rm max}$, where $\lambda_i^{\rm max}$ is determined by $\beta(\lambda_i^{\rm max}) = 1$ even when all other couplings are set to zero. Hence, when any ${\bar \lambda}_i = 1$, we expect that perturbation theory is no longer reliable. The dashed horizontal line at 1 is to guide the eye to the approximate perturbativity boundary. For $N_S = 2$ this happens almost instantaneously, while for $N_2 = 3$ it happens at about 1.4 TeV.}
\label{fig:scalarrge}
\end{figure}%

We plot some examples of RGE evolution for the normalized couplings in Figure~\ref{fig:scalarrge}. We begin the evolution by fixing $\lambda_H$ from the Higgs mass, $\lambda_{\phi S} N_S = 4.3$ to fit the gamma-ray line, $\lambda_S = 9.3/N_S^2$ for vacuum stability, $\lambda_{HS} N_S = -2.2$ for a 50\% enhancement of $h \to \gamma\gamma$, $y_t$ from the top mass, and $\lambda_{\phi H} = \lambda_\phi = 0$ at low scales. The result is that, for $N_S = 2$, perturbativity is lost almost immediately on evolving to higher scales, while for $N_S = 3$ it is lost around the TeV scale. This suggests that the most reasonable interpretation of these models is as composite models, where the scalars are bound states.~\footnote{Note also that, since $\lambda_{\phi S}$ is so large, this statement applies even if we consider smaller Higgs diphoton enhancements; in other words, the Fermi line alone suggests that we are considering a composite model.} This makes large couplings $\lambda_S$ and $\lambda_{\phi S}$ seem more natural; however, the large value of $\lambda_{HS}$ suggests that perhaps the Higgs would be composite too. In this case, the small values of $\lambda_{\phi H}$ (required by direct and indirect detection bounds) and $\lambda_H$ (required by the Higgs mass measurement) seem hard to reconcile with the idea that composite states would generically be strongly coupled to each other. We could move to larger values of $N_S$ to postpone the loss of perturbativity to higher energy scales, but would still face a puzzle in the small value of $\lambda_{\phi H}$. For instance, by choosing $N_S = 6$, we can postpone the loss of perturbativity to a scale of 250 TeV. Then, Eq.~\ref{eq:lambdaphiHestimate} has a factor of $\frac{1}{N_S} \log\frac{\Lambda}{m_S}$; we have increased both the log and $N_S$, and so have not really helped solve the problem. Indeed, solving the RGE shows that $\lambda_{\phi S}$ blows up first; the RGE $16\pi^2 \beta(\lambda_{\phi S}) = 8 \lambda_{\phi S}^2$ tells us that the coupling will blow up at about
\beq 
\log\frac{\Lambda_{\rm UV}}{m_S} \approx \left.\frac{2\pi^2}{\lambda_{\phi S}}\right|_{\mu = m_S} \approx \frac{2 \pi^2 N_S}{4.3}.
\eeq
Thus, the factor of $\frac{1}{N_S} \log\frac{\Lambda_{\rm UV}}{m_S}$ in Eq.~\ref{eq:lambdaphiHestimate} does not scale with $N_S$, and the problem is unavoidable:  generically, RG evolution is in conflict with the lack of a direct detection signal. In short, although it is very appealing to consider the idea that new charged particles explain both dark matter annihilation and the $h \to \gamma\gamma$ enhancement, in this simplest model our closer look has shown that quantum corrections spoil the nice idea.

Aside from raising $N_S$, one could try to ameliorate the Landau problem by making the electric charge of $S$ larger. For instance, making $S$ charge 2 instead of 1 will reduce the required values of $\lambda_{\phi S}$ and $\lambda_{HS}$ by a factor of 4. This brings the loop induced $\lambda_{\phi H}$ to about the level of the XENON bound provided the cutoff scale remains at about 1 TeV. Note that $S$, and any other new charged particles we discuss in this paper, must decay promptly to avoid strong collider bounds on heavy stable charged particles (HSCPs). At large charges of $S$, any effective operator allowing it to decay to charged standard model particles will be of high dimension, so that larger charges rapidly lead to longer lifetime decays (or require extending the model with a bevy of new intermediate states of progressively smaller charges). A detailed analysis of $S$ decays and collider constraints is beyond the scope of this paper, but we expect that our conclusions can be weakened only mildly by constructing models where $S$ has larger charge and has escaped detection so far. There is a loophole in this statement. Particles in new SU(2)$_L$ multiplets naturally present precisely a set of particles of progressively smaller charge, and also automatically come with decay channels: the particle of charge $n+1$ can decay to its SU(2)$_L$ partner of charge $n$, together with an off-shell $W$ boson. Such decays could have a detectably long lifetime, but the charged particles are not typically collider-stable. Our statement that operators of high dimension are required for the decay is still true for the decay of the neutral state in the multiplet, but the neutral state is not subject to HSCP bounds. Such models have been studied recently in Ref.~\cite{Kopp:2013mi}, to which we refer the reader for a more detailed discussion.

Finally, we comment that the analysis of Ref.~\cite{Baek:2012ub} is rather similar to ours, including also RG effects and observing the existence of a low-scale Landau pole. However, one of our central conclusions, that attempts to fit the Fermi-LAT line and the $h \to \gamma\gamma$ rate are necessarily in tension with direct detection bounds, is less clear in their analysis. Their discussion of thermal relic abundance, which we have not considered here, is interesting.

\subsection{Resonant annihilation}
\label{sec:resonantannihilation}

A scenario that has received a great deal of attention as a possible explanation of the Fermi-LAT gamma line is resonant annihilation~\cite{Buckley:2012ws, Lee:2012bq, Das:2012ys, Tulin:2012uq, Dudas:2012pb,Lee:2012wz,SchmidtHoberg:2012ip,Chalons:2012xf,Bai:2012yq}. By exploiting a pole in a propagator to enhance the annihilation cross section, the large couplings and nearby Landau poles we encountered in the previous subsection may be avoided. The cost is tuning the mass of an intermediate particle to be close to twice the dark matter mass, for no deep reason. We will consider primarily the case of Majorana fermion dark matter in this subsection. Two identical Majorana fermions, annihilating at low velocities, form a CP-odd initial state, so the intermediate particle should be a pseudoscalar rather than a scalar. (In other words, an intermediate scalar would lead to a suppressed $p$-wave, rather than $s$-wave, annihilation cross section.) This might seem to preclude the generation of Higgs-related signals, since the Higgs does not mix with pseudoscalars in the absence of CP violation. However, any new charged particles that affect the $h \to \gamma\gamma$ rate would, in general, allow the presence of new CP-violating phases, in the presence of which a direct detection signal could be generated. Alternatively, although we will not discuss it in detail, one could consider Dirac fermion dark matter, which could annihilate through a CP-even scalar rather than a pseudoscalar.

\subsubsection{$s$-channel exchange of a boson}

\begin{figure}[!h]\begin{center}
\includegraphics[width=0.4\textwidth]{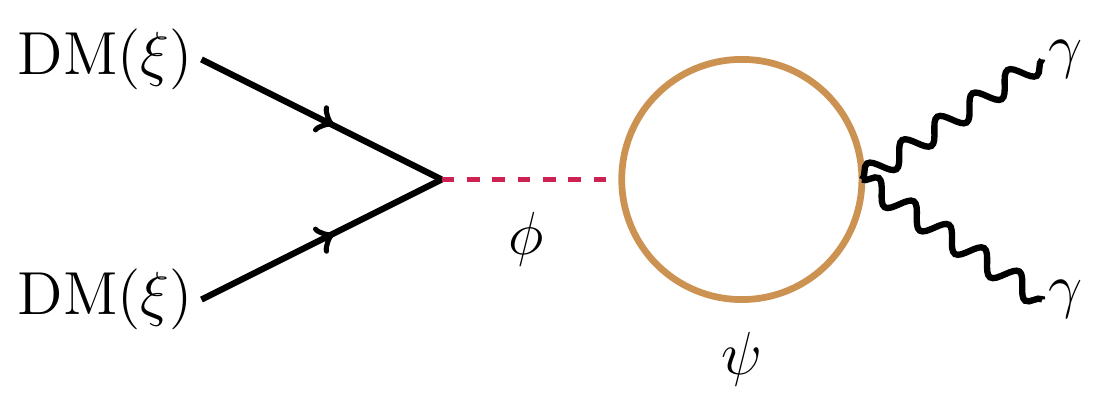} 
\end{center}
\caption{The annihilation processes leading to two photons in resonant annihilation model.}
\label{fig:process2}
\end{figure}

We will take DM to be a Majorana fermion $\xi$ with mass $m_\xi$, annihilating through an intermediate pseudoscalar boson $\phi$ with mass $m_\phi$ and width $\Gamma_\phi$. The intermediate state $\phi$ couples to photons through $N_\psi$ species of charged light fermions $\psi$ and $\psi^c$ (with conjugate charges) with mass $m_\psi$ and charge $Q_\psi$. The Lagrangian is given by 
\beq
m_\psi \psi^c \psi + \frac{1}{2} m_\xi \xi \xi + \frac{i}{2} g_\xi \xi \xi \phi + i g_\psi \phi \psi^c \psi + {\rm h.c.}
\eeq
(with $g_\xi, g_\psi$ real couplings and $\phi$ one real degree of freedom) leading to the annihilation process as depicted in Fig.~\ref{fig:process2}. We will take the dark matter mass $m_\xi = 130$ GeV.

\begin{figure}[!h]\begin{center}
\includegraphics[width=0.95\textwidth]{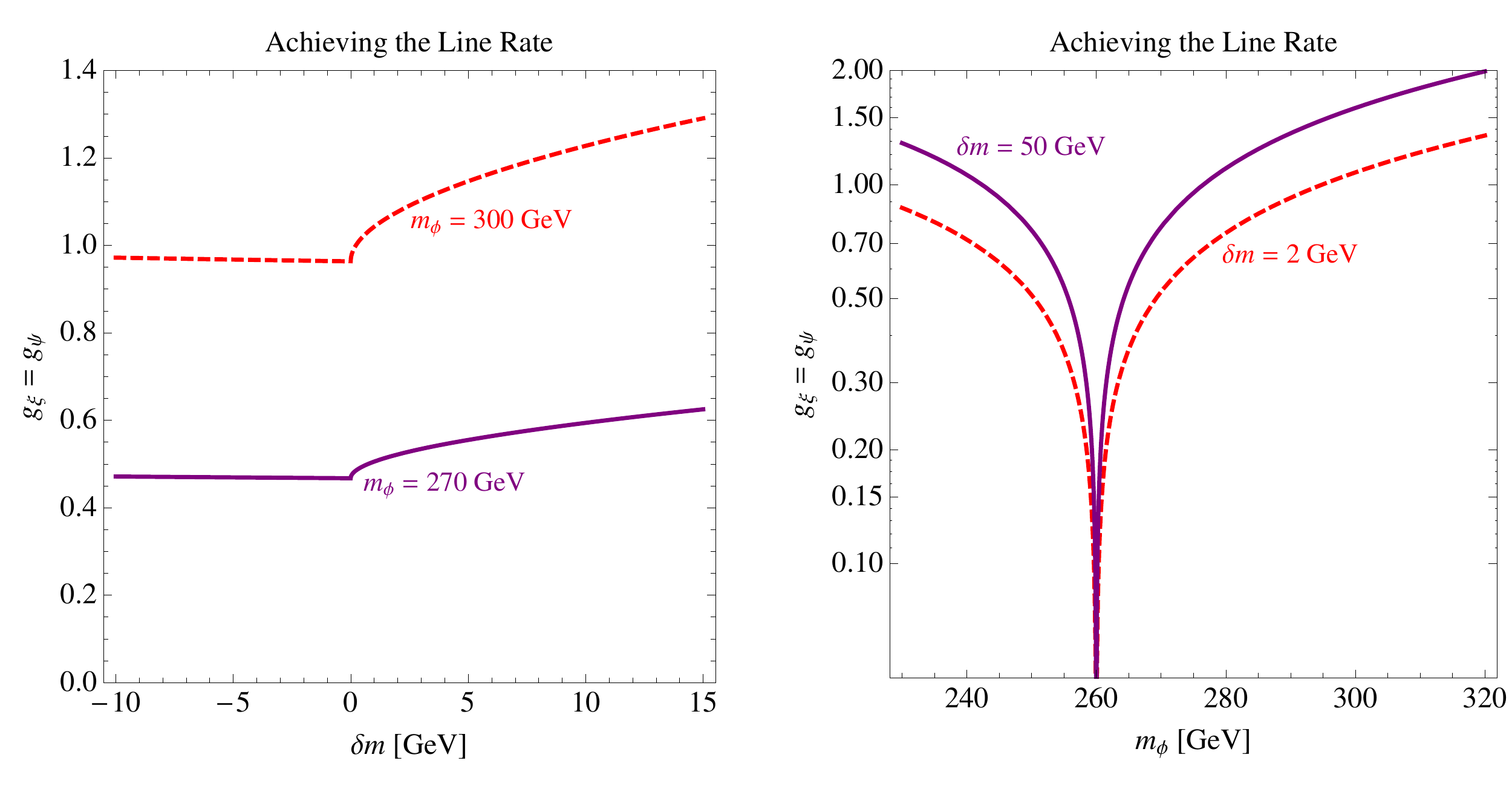}
\end{center}
\caption{Resonant annihilation. Left: contours of $(\sigma( \xi \xi \to \gamma \gamma) v) = 10^{-27}$ cm$^{3}$s$^{-1}$ in the ($\delta m, g_\xi = g_\psi$) plane, where $\delta m \equiv m_\psi - m_\xi$. Purple, solid: $m_\phi = 270$ GeV; red, dashed: $m_\phi=300$ GeV. At right: the same contours as a function of $m_\phi$; purple: $\delta m = 50$ GeV; red, dashed: $\delta m = 2$ GeV.}
\label{fig:fermion}
\end{figure}

The formula of the cross section for $\xi \xi \to \gamma\gamma$, allowing for multiple fermionic states to run in the loop, is given in~\cite{Bergstrom:1997fh,Bern:1997ng,Tulin:2012uq}:
\beq
\sigma v = \frac{\alpha^2}{4\pi^3} \frac{g_\xi^2}{\left(4 m_\xi^2 - m_\phi^2\right)^2 + m_\phi^2 \Gamma_\phi^2} \left|\sum_\psi N_\psi Q_\psi^2 g_\psi m_\psi I\left(\frac{m_\psi^2}{m_\xi^2}\right)\right|^2,
\eeq
where for $m_\psi > m_\xi$, $I\left(\frac{m_\psi^2}{m_\xi^2}\right) = -\left({\rm arctan} \sqrt{\frac{1}{m_\psi^2/m_\xi^2-1}}\right)^2$.
Demanding $(\sigma v)( \xi\xi \to \gamma \gamma) = 10^{-27}$ cm$^{3}$s$^{-1}$, and assuming all the fermions in the loop are degenerate, one obtains $g_\xi g_\psi N_\psi Q_\psi^2$ as a function of $\delta_m\equiv m_\psi-m_\xi$. We determine $\Gamma_\phi$ by summing the partial widths $\phi \to \xi \xi$, $\phi \to \psi \psi$, and $\phi \to \gamma\gamma$ (when the modes are kinematically accessible) using the formulas in Ref.~\cite{Tulin:2012uq}. Because the partial widths depend on $g_\xi$, $g_\psi$, and the charge and species count of $\psi$ independently, we plot only the case $g_\xi = g_\psi, N_\psi = Q_\psi = 1$. The result is presented in Fig.~\ref{fig:fermion}. We see that the $\psi$ particles should not be much heavier than $\xi$ if we want to fit the line without large couplings. More importantly, we require an approximate resonance condition, $m_\phi \approx 2 m_\xi$. With a high degree of fine-tuning the couplings can be made quite small; e.g., we can achieve $g_\xi = g_\psi \approx 0.1$ for $\left|m_\phi - 2 m_\xi\right| \approx 0.5$ GeV.  With less extreme fine-tuning, e.g, a 10\% coincidence in masses, the required couplings are approximately 1.

If we also want to alter the $h \to \gamma\gamma$ rate through loops of this charged particle, we need a more complicated structure. Following Ref.~\cite{ArkaniHamed:2012kq}, we assume that $\psi$ carries only hypercharge while another field $\chi$ carries $SU(2)_W$ charge, $\psi,\psi^c\sim(1,1)_{\mp 1},\;\;\chi,\chi^c\sim(1,2)_{\pm \frac{1}{2}}.$
The Lagrangian is:
\ba\label{eq:doubL}-\mathcal{L}=m_\psi\psi\psi^c+m_\chi\chi\chi^c+yH\psi\chi+y^cH^\dag\psi^c\chi^c+cc.\ea
and we also add, for generality, a coupling $g_\chi \phi \chi^c \chi$. We will be interested in the general case with a non-zero CP violating phase $\arg(y y^c m_\psi^* m_\chi^*) \neq 0$. At loop level, this will also force us to consider complex values of the couplings $g_\psi$ and $g_\chi$. In this case, the one-loop RGEs for our new Yukawa couplings (neglecting SM gauge coupling effects) are extracted from the general formulas~\cite{Machacek:1983tz,Machacek:1983fi,Machacek:1984zw,Luo:2002ti}:
\beq
16\pi^2 \beta(g_\xi) &=& g_\xi \left(2 \left|g_\xi\right|^2 + \left|g_\psi\right|^2 + 4 \left|g_\chi\right|^2\right) \\
16\pi^2 \beta(g_\psi) &=& g_\psi \left(3 \left|g_\psi\right|^2 + \left|g_\xi\right|^2 + 4 \left|g_\chi\right|^2 + \left|y\right|^2 + \left|y^c\right|^2\right) - 4 g_\chi^\dagger y y^c \\
16\pi^2 \beta(g_\chi) &=& g_\chi \left(5 \left|g_\chi\right|^2 + \left|g_\xi\right|^2 + 2 \left|g_\psi\right|^2 + \frac{1}{2} \left(\left|y\right|^2 + \left|y^c\right|^2\right)\right) - 2 g_\psi^\dagger y y^c \\
16\pi^2 \beta(y) &=& \frac{1}{2} y \left(\left|g_\chi\right|^2 + \left|g_\psi\right|^2 + 5 \left|y\right|^2 + 2 \left|y^c\right|^2\right) - 2 y^{c\dagger} g_\psi g_\chi \\
16\pi^2 \beta(y^c) &=& \frac{1}{2} y^c \left(\left|g_\chi\right|^2 + \left|g_\psi\right|^2 + 5 \left|y^c\right|^2 + 2 \left|y\right|^2\right) - 2 y^\dagger g_\psi g_\chi
\eeq
We will evolve these RGEs to examine the extent to which annihilating through resonant enhancement relieves the Landau pole problem of the all-scalar model.

The modified $h \to \gamma\gamma$ rate, in the limit of large $m_{\psi,\chi}$, is determined by the low energy theorems to be~\cite{Shifman:1979eb,ArkaniHamed:2012kq,Voloshin:2012tv,McKeen:2012av}:
\beq
\mu_{\gamma\gamma} =  \left|1 + \frac{1}{{\cal A}_{\rm SM}}\frac{2}{3} Q_\psi^2 \frac{\partial}{\partial\log v}\log \det {\cal M}^\dagger{\cal M} \right|^2 + \left|\frac{2}{{\cal A}_{\rm SM}} Q_\psi^2 \frac{\partial}{\partial\log v}\arg \det {\cal M}\right|^2,
\eeq
where the first term arises from the modified CP-even $hF_{\mu\nu}F^{\mu\nu}$ vertex and the second term from the CP-odd $hF_{\mu\nu}{\tilde F}^{\mu\nu}$ term. Here ${\cal A}_{\rm SM} = -6.49$ represents the SM amplitude. As in the scalar case, the result is modified by familiar loop functions that correct for finite mass, which are given in the CP-even and CP-odd case respectively by~\cite{Djouadi:2005gj}
\beq
A_{1/2}(\tau_f)  &=& \frac{3}{2} \tau_f \left(1 + (1-\tau_f) \arcsin^2\sqrt\frac{1}{\tau_f}\right)\\
{\tilde A}_{1/2}(\tau_f) &=& \tau_f \arcsin^2\sqrt\frac{1}{\tau_f},
\eeq
(with $\tau_f = 4 m_f^2/m_h^2$) and asymptote to 1 when $2 m_f \gg m_h$. (Note that there is a minor error in Ref.~\cite{ArkaniHamed:2004yi}, which assumes the same loop function for the scalar and pseudoscalar decay modes.)

\begin{figure}[!h]\begin{center}
\includegraphics[width=\textwidth]{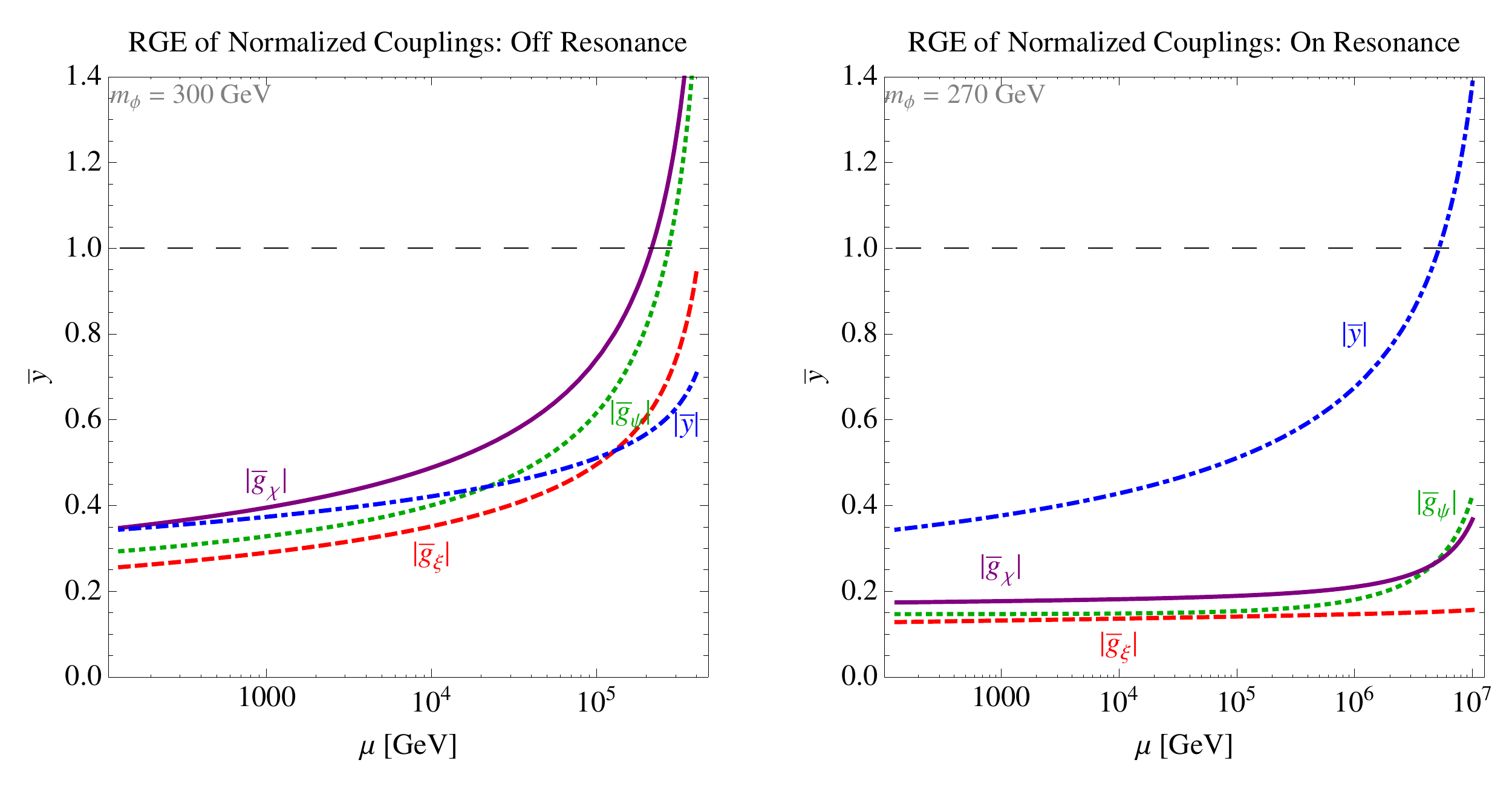}
\end{center}
\caption{Renormalization group evolution of ``normalized couplings'' for the case of annihilation through a scalar resonance. Parameters are as in the ``moderate phase'' case described in the text; the ``small phase'' case has similar RG evolution. At left, we have chosen $m_\phi \approx 300$ GeV, far enough off-resonance that the Fermi-LAT line rate requires $g\equiv g_\xi = g_\psi = g_\chi \approx 1.1$. In this case the couplings become non-perturbative at scales of order 100s of TeV. At right, we show results for $m_\phi \approx 270$ GeV, closer to the resonance with $g \approx 0.55$. This significantly increases the perturbativity range for the couplings $g$. Thus, unlike the all-scalar model for the gamma ray lines, in this scenario the larger loop contributions arise from fitting the $h \to \gamma\gamma$ rate, and in fact the Higgs quartic running (not shown here) is the most important effect that will require new physics at low scales~\cite{ArkaniHamed:2012kq}.}
\label{fig:scalarrge}
\end{figure}%

We choose two representative points in parameter space that fit a 50\% enhanced $h \to \gamma\gamma$ rate. In both cases, we arrange for the light mass eigenstate to be at 140 GeV, such that it is near the DM mass but slightly too heavy for the dark matter to annihilate into two of our new charged fermions.
\begin{itemize}
\item Moderate phase: $m_\psi = m_\chi = 346$ GeV, $y = 1.37 e^{0.2 \pi i}$, $y^c = 1.37$. The mass eigenvalues are 140 and 577 GeV. As we will see in the next section, this point is excluded by the nonobservation of an electron EDM: it predicts $d_e/e = 9.3 \times 10^{-27}~{\rm cm}$. However, this exclusion can be avoided if the EDM is canceled by another contribution which must be tuned to about the 10\% level. (And note that our phase is already somewhat small; maximal CP violation would require the EDM to be tuned at a few-percent level.)
\item Small phase: $m_\psi = m_\chi = 333$ GeV, $y = 1.11 e^{0.02 \pi i}$, $y^c = 1.11$. The mass eigenvalues are 140 and 526 GeV. This point is currently safe from EDM constraints, predicting $d_e/e = 7.3 \times 10^{-28}~{\rm cm}$, but would likely be detected with next-generation electron EDM measurements~\cite{Vutha:2009ux}.
\end{itemize}
In both cases, we will also consider equal couplings $g \equiv g_\xi = g_\psi = g_\chi$ of the pseudoscalar resonance, chosen to achieve a gamma-ray line rate of $1.0 \times 10^{-27}~{\rm cm}^3/{\rm s}$. These turn out to be relatively insensitive to the mass of the heavy eigenstate. We consider two choices of resonance mass: $m_\phi = 270$ GeV, which requires $g \approx 0.55$, and $m_\phi = 300$ GeV, which requires $g \approx 1.1$. The renormalization group evolution of these couplings is plotted in Figure~\ref{fig:scalarrge}. The main qualitative feature to note is that, compared to the model in Section~\ref{sec:allscalar} without resonant enhancement, the couplings stay perturbative until much higher scales. There is another crucial RG effect not visible in our plots, discussed in detail in Ref.~\cite{ArkaniHamed:2012kq}: the large Yukawa couplings $y$ and $y^c$ will drive the Higgs quartic coupling negative at low scales. This vacuum instability suggests that the model must be altered, for instance by adding light superpartners of the $\psi$ and $\chi$ fields to cancel their effect on the Higgs quartic. However, we should emphasize that if one only wishes to explain a gamma ray line signal, this concern is not relevant, and the couplings $g_{\xi,\psi}$ involved in dark matter annihilation can remain perturbative to very high energies if the $\phi$ mass is tuned so that annihilation is on resonance.

\subsubsection{Direct detection}

\begin{figure}[!h]\begin{center}
\includegraphics[width=0.3\textwidth]{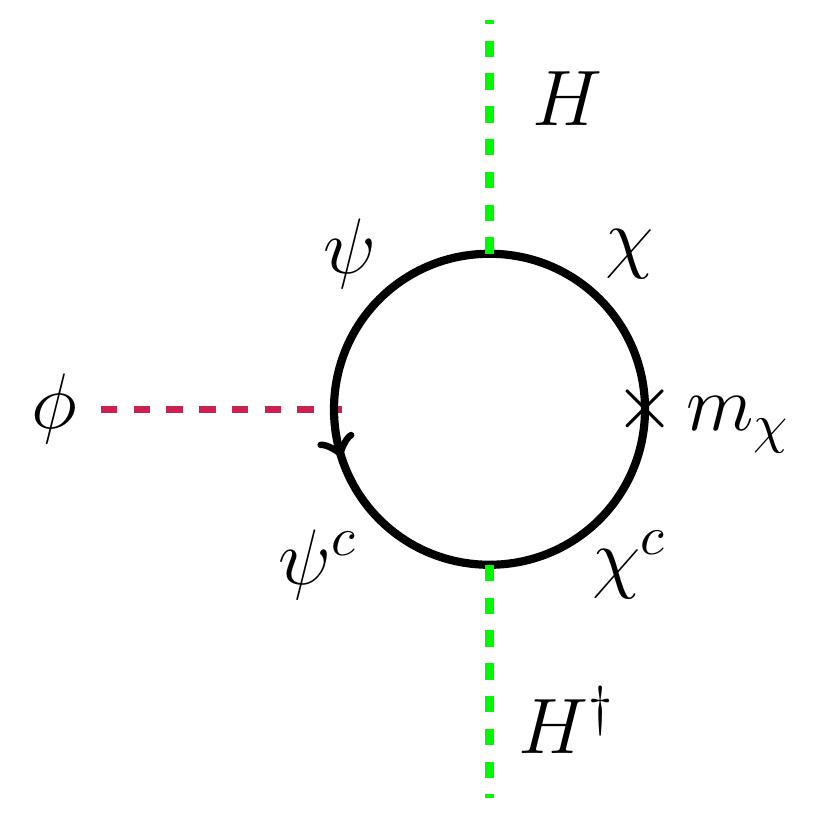}
\end{center}
\caption{The induced Higgs/pseudoscalar mixing through a CP-violating loop of $\psi$ and $\chi$ fermions.}
\label{fig:CWloop}
\end{figure}%

 In the presence of CP violation, scalars and pseudoscalars can generically mix. Integrating out $\psi$ and $\chi$, we find that the Coleman-Weinberg potential contains a term, generated by the loop in Figure~\ref{fig:CWloop}:
\beq
V_{\rm CW} \supset \frac{{\rm Im}(y y^c)}{4\pi^2}\left(g_\chi m_\psi + g_\psi m_\chi\right)\left(1 + \frac{m_\psi^2}{m_\chi^2-m_\psi^2}\log\frac{m_\psi^2}{\mu^2} - \frac{m_\chi^2}{m_\chi^2 - m_\psi^2}\log\frac{m_\chi^2}{\mu^2}\right)H^\dagger H \phi.
\eeq
when expanding around the origin. Here we have taken the mass terms $m_\psi, m_\chi$ to be real, so that the phases would be contained solely in the Yukawas $y, y^c$. A nontrivial phase can lead to significant mixing. To calculate the mixing, we expand the Coleman-Weinberg potential around the physical Higgs VEV and evaluate at a renormalization scale $\mu = \sqrt{m_\phi m_h}$. We tabulate the resulting mixing angles between Higgs and $\phi$, $\theta_{h\phi}$, for four cases (moderate or small phase, on or off resonance) in Table~\ref{tab:higgsDMcouplings}. 

The mixing between $\phi$ and higgs lead to an effective operator $\bar{D} \gamma_5 D \bar{q}q$, with the Dirac fermion $D=(\xi, i\sigma_2\xi^*)^T$ and $q$ the SM quarks, relevant for direct detection. In the non-relativistic limit, one could see that the scattering rate off nucleons through this operator is suppressed by the momentum transfer (the amplitude is $\propto \vec{s}_\xi \cdot \vec{q}$), leading to an interesting recoil spectrum rising at high recoil energy~\cite{Fan:2010gt}. However, the small rate limits its possibility of being detected. The current XENON direct detection limit is $g_\xi\theta_{h\phi} \lesssim 15(m_\phi/300\,{\rm GeV})^2$, about two orders of magnitude above $g_\xi\theta_{h\phi} \approx 0.1$ in the case with a moderate CP phase. Similarly, the conjunction of pseudoscalar and scalar couplings leads to a $p$-wave suppressed indirect detection cross section for annihilation through the Higgs/$\phi$ mixture. Thus, the pseudoscalar resonance models are safely out of reach from direct and indirect detection bounds from loop induced DM-Higgs couplings for the foreseeable future.

\begin{table}[h]
\begin{center}
\begin{tabular}{l|ll}
$\theta_{h\phi}$& Phase $0.2\pi$ & Phase $0.02\pi$ \\
\hline
$m_\phi = 270$ GeV & 0.076 & 0.0047 \\
$m_\phi = 300$ GeV & 0.11 & 0.0068
\end{tabular}
\caption{Mixing angles between Higgs and the pseudo-scalar $\phi$, for both moderate and small phase and on- or off-resonance. }
\label{tab:higgsDMcouplings}
\end{center}
\end{table}

In this subsection, all of our numerical choices have fixed a 50\% enhancement of the $h \to \gamma\gamma$ rate. Given the latest CMS results, this is likely to be an overestimate of any real effect (though it is still below the ATLAS central value). Because the conclusion has been that, in the resonant models, there is no tight connection between fitting the Fermi line and increasing $h \to \gamma\gamma$ (since the direct detection loop is unconstrained), any smaller $h\to\gamma\gamma$ deviation would be even safer, and easily accommodated in resonant models of the Fermi line.

\section{Correlation between CP-odd and CP-even observables}
\label{sec:CPodd}

\subsection{Operators and corresponding observables}
\label{sec:ops}
We will assume that new charged matter does not interact or mix with the SM fermions at tree level and they do not contribute to the EDMs of the SM fermions at one-loop order. (Thus, we will also not discuss the discrepancy in the measured muon $g-2$, which would typically require new physics with leptonic quantum numbers exerting a one-loop effect.) The one-loop EDM is generically ruled out unless the new CP violating phases are tuned to be small ($\lsim 10^{-2}$). However, new charged particles could still contribute at the two-loop order to the SM fermion EDMs through Barr-Zee type diagrams~\cite{Barr:1990vd}. To see the correlations between the CP-odd observables (including EDMs) and CP-even observables, it will be useful to perform an operator analysis first, which strictly speaking is only valid in the limit when the charged matter is heavy and could be integrated out.  We will use operator analysis to clarify the correlations of the observables, whereas for the numerical evaluations we will use the full-fledged loop calculations. Charged matter with physical phases contributes to 6 CP violating dimension-six operators built out of the Higgs and the SM gauge fields. Among them, two involve the $SU(3)$ color field strength, one of which is the famous Weinberg operator~\cite{Weinberg:1979sa}. We only inspect the four operators generated by loops of colorless particles. They and their corresponding CP-even operators with similar structures are, following the notations in~\cite{Buchmuller:1985jz},
\beq
{\cal O}_W&=&\epsilon_{abc}W^{a \nu}_\mu W^{b \lambda}_\nu W^{c \mu}_\lambda, \quad {\cal O}_{\tilde{W}}=\epsilon_{abc}\tilde{W}^{a \nu}_\mu W^{b \lambda}_\nu W^{c \mu}_\lambda \nonumber \\
{\cal O}_{hW}&=&H^\dagger H W^{a}_{ \mu\nu} W^{a \mu\nu},  \quad \quad{\cal O}_{h\tilde{W}}=H^\dagger H \tilde{W}^{a}_{  \mu\nu} W^{a \mu\nu} \nonumber  \\
{\cal O}_{hB}&=&H^\dagger H B_{\mu\nu} B^{ \mu\nu},  \quad \quad \quad{\cal O}_{h\tilde{B}}=H^\dagger H \tilde{B}_{\mu\nu} B^{ \mu\nu} \nonumber  \\ 
{\cal O}_{WB}&=&(H^\dagger \sigma^a H) W^{a}_{ \mu\nu} B^{ \mu\nu},  \quad {\cal O}_{\tilde{W}B}=(H^\dagger \sigma^a H) \tilde{W}^{a}_{ \mu\nu} B^{ \mu\nu},
\label{eq:ops}
\eeq 
where $\sigma^a$ denotes the three Pauli matrices and $a$ is the isospin index. The operators have coefficients $a_i$ bounded by the interval $[-1/ \Lambda_{\rm neg}^2, 1/\Lambda_{\rm pos}^2]$, where $\Lambda$ is some high scale.\footnote{In some literature~\cite{Giudice:2005rz, Jung:2008it}, more CP-odd operators were listed. As we show in Appendix~\ref{app:CPoddops}, those additional operators can be written in terms of the operators in Eq.~\ref{eq:ops} using the equations of motion.}

Now we specify the observables these operators contribute to. 
It is well known that ${\cal O}_{WB}$ gives the $S$ parameter in the electroweak precision tests (EWPT),
 \beq
 S= \frac{4s_Wc_Wv^2a_{WB}}{\alpha}, \nonumber
 \eeq
 where $s_W \equiv \sin \theta_W$, $c_W \equiv \cos \theta_W$, and $v=246$ GeV. 
 
 Among all the operators, ${\cal O}_W, {\cal O}_{WB}$ (${\cal O}_{\tilde{W}}, {\cal O}_{h\tilde{W}},  {\cal O}_{\tilde{W}B}$) modify CP-even TGCs (CP-odd TCGs). More concretely, the general triple gauge couplings could be parametrized as~\cite{Hagiwara:1986vm},
\beq
{\cal L}_{WWV}/g_{WWV}&=&i g_1^V\left(W^+_{\mu\nu}W^{-\mu} V^\nu-h.c.\right)+i \kappa_V W_\mu^+ W_\nu^- V^{\mu\nu}+\frac{i\lambda_V}{m_W^2}W^+_{\mu\nu}W^{-\nu\lambda}V_{\lambda}^\mu \nonumber \\
&&+i\tilde{\kappa}_VW_\mu^+ W^-_\nu \tilde{V}^{\mu\nu}+\frac{i\tilde{\lambda}_V}{m_W^2}W^+_{\mu\nu}W^{-\nu\lambda}\tilde{V}_{\lambda}^\mu +\cdots,
\label{eq:tgc}
\eeq
where $V$ is either $Z$ or $\gamma$ and $g_{WWZ} =- e \cot \theta_W, g_{WW\gamma}=-e$. The first line of Eq.~\ref{eq:tgc} contains CP-even TGCs while the second line contains CP-odd TGCs. The dots represent C-violating TGCs arising from operators at high orders in the SM effective theory. In the SM, $g_1^V=1$, $\kappa_V=1$ and $\lambda_V=0$ at tree level while $\tilde{\kappa}_V$ and $\tilde{\lambda}_V$ are zero even at one-loop order in the SM due to unitarity of the CKM matrix. The contributions to the parameters in Eq.~\ref{eq:tgc} from high-dimensional operators are
\beq
\delta \kappa_Z &=& \frac{v^2s_W}{c_W}a_{WB},\quad \quad\delta \kappa_\gamma =-  \frac{v^2c_W}{s_W}a_{WB},\nonumber \\
\delta \lambda_Z &=& \delta \lambda_\gamma= \frac{6m_W^2a_W}{g}, \nonumber \\
\delta \tilde{\kappa}_Z &=&2v^2a_{h\tilde{W}}+\frac{v^2s_W}{c_W}a_{\tilde{W}B},\quad \quad \delta \tilde{\kappa}_\gamma =2 v^2a_{h\tilde{W}}-  \frac{v^2c_W}{s_W}a_{\tilde{W}B},\nonumber  \\
\delta\tilde{ \lambda}_Z &=& \delta \tilde{ \lambda}_\gamma= \frac{6m_W^2a_{\tilde{W}}}{g}. 
\eeq

The operators ${\cal O}_{hW}, {\cal O}_{hB}, {\cal O}_{WB}, {\cal O}_{h\tilde{W}}, {\cal O}_{h\tilde{B}}, {\cal O}_{\tilde{W}B}$ also modify the Higgs decays
\beq
\mu_{\gamma\gamma}&\equiv&\frac{\Gamma(h\to \gamma\gamma)}{\Gamma(h\to \gamma\gamma)_{\rm SM}} \nonumber \\
&=&\left|1+\frac{8\pi v^2(s_W^2 a_{hW}+c_W^2a_{hB}-s_Wc_Wa_{WB})}{\alpha {\cal A}_{\rm SM}}\right|^2+\left|\frac{8\pi v^2(s_W^2 a_{h\tilde{W}}+c_W^2a_{h\tilde{B}}-s_Wc_Wa_{\tilde{W}B})}{\alpha {\cal A}_{\rm SM}}\right|^2\quad
\label{eq:decay}
\eeq
where ${\cal A}_{\rm SM}={\cal A}_{W}+{\cal A}_t\approx -6.5$ is proportional to the SM amplitude.

The four CP-odd operators also contribute to the electron EDM at the one-loop order through Barr-Zee type diagrams~\cite{Barr:1990vd},
\beq
\frac{d_f}{e}& = &\frac{Q_fm_f}{4\pi^2}\Biggl(s_W^2a_{h\tilde{W}}\ln\left(\frac{m_h^2}{\Lambda_{h\tilde{W}}^2}\right)+c_W^2a_{h\tilde{B}}\ln\left(\frac{m_h^2}{\Lambda_{h\tilde{B}}^2}\right)-s_Wc_Wa_{\tilde{W}B}\ln\left(\frac{m_h^2}{\Lambda_{\tilde{W}B}^2}\right)\nonumber\\
& & \quad -\frac{3T_3e}{2Q_fs_W}a_{\tilde{W}}\ln\left(\frac{m_h^2}{\Lambda_{\tilde{W}}^2}\right)\Biggr),
\eeq
where $T_3$ is the isospin of the charged matter.

Now we review the experimental status of the observables. Currently the most stringent constraints on CP-even TGCs are from the measurement of differential cross sections of $e^+e^- \to W^+W^-$ at LEP 2~\cite{Achard:2004zw} and those of $pp \to W^+W^-/WZ\to l\nu jj$ at CMS~\cite{Chatrchyan:2012bd}: these lead to constraints $-0.038 <\lambda_{Z,\gamma}<0.03$, $-0.11<\Delta \kappa_\gamma < 0.14$ at 95\% CL. ATLAS also measures CP-even TGCs in the fully leptonic channel, obtaining weaker constraints due to the smaller branching fractions~\cite{:2012me}. The CP-odd TGCs are also studied at LEP and the Tevatron~\cite{Abdallah:2008sf, Abachi:1996hw}: $\tilde{\kappa}_Z=-0.09^{+0.08}_{-0.05}, \tilde{\lambda}_Z=-0.08\pm0.07, |\tilde{\lambda}_\gamma| \leq 0.32$ at 95\%. The Higgs data start to constrain the modifications of Higgs couplings, though due to the large uncertainties the constraints are weak at the moment. Nonetheless, it is very interesting that $O_{hW}$ and $O_{hB}$ are becoming as important as the traditional electroweak precision operators. For EDM experiments, the current electron EDM bound is $d_e/e < 1.05 \times 10^{-27}$ cm (90\%)~\cite{Hudson:2011zz,Kara:2012ay} and the neutron EDM is also stringently constrained: $d_n/e < 2.9 \times 10^{-26}$ cm (90\%)~\cite{Baker:2006ts}. (A bound on the EDM of mercury, $d(^{199}{\rm Hg})/e < 3.1 \times 10^{-29}$ cm (95\%)~\cite{Griffith:2009zz}, is also noteworthy.) It is expected that there will be an update in the electron EDM measurement in the near future, improving the current bound by an order of magnitude~\cite{Vutha:2009ux}. We summarize the current experimental constraints on the coefficients of these operators in Table~\ref{tab:constraints}.\footnote{One should be careful in interpreting the bounds. For example, ${\cal O}_{hW}, {\cal O}_{hB}$ are always generated with a coefficient $\alpha_{EM}/\pi \sim 10^{-3}$. Thus Higgs data itself does not put a strong constraint on the mass of the charged particle so far. }

\begin{table}[h]
\begin{center}
\begin{tabular}{|c|c|c|}
\hline
${\cal O}$ & Experiments & $\Lambda_i$(TeV) \\
\hline
${\cal O}_{WB}$ & EWPT~\cite{Beringer:1900zz} & 12.6 (90\%)~\cite{Han:2008es} \\
${\cal O}_{hW}, {\cal O}_{hB}$ & $h \to \gamma\gamma$~\cite{CMSdiphotonupdate, ATLASdiphotonupdate} &  1.8 (68\%)/ 3.3 (68\%)\\
${\cal O}_W$ & CP-even TGCs~\cite{Achard:2004zw, Chatrchyan:2012bd} & 1.3 (95\%) \\
\hline
${\cal O}_{\tilde{W}}$ & CP-odd TGCs~\cite{Abdallah:2008sf, Abachi:1996hw}/electron EDM~\cite{Hudson:2011zz,Kara:2012ay}& 0.5 (95\%)/ 38 (90\%) \\
${\cal O}_{h\tilde{W}}$ &CP-odd TGCs~\cite{Abdallah:2008sf, Abachi:1996hw}/electron EDM~\cite{Hudson:2011zz,Kara:2012ay} & 0.9 (95\%)/ 24 (90\%)\\
${\cal O}_{h\tilde{B}}$ &electron EDM~\cite{Hudson:2011zz,Kara:2012ay} & 48 (90\%) \\
${\cal O}_{\tilde{W}B}$ &CP-odd TGCs~\cite{Abdallah:2008sf, Abachi:1996hw}/electron EDM~\cite{Hudson:2011zz,Kara:2012ay} & 0.5 (95\%)/ 35 (90\%)\\
\hline
\end{tabular}
\caption{Current experimental bounds on operator coefficients (the CL are in the parenthesis). The operator coefficient $a_i$ is bounded by the interval  $[-1/ \Lambda_{\rm neg}^2, 1/\Lambda_{\rm pos}^2]$. The $\Lambda_i$ (in TeV) shown in the table is the average of $\Lambda_{\rm neg}$ and $\Lambda_{\rm pos}$. When $a_i$ is experimentally bounded to be negative (positive) definite, we only quote $\Lambda_{\rm neg}$ ($\Lambda_{\rm pos}$). For constraints from $h\to\gamma\gamma$, we fix Higgs mass at 125 GeV.}
\label{tab:constraints}
\end{center}
\end{table}

\subsection{Correlation between CP-odd and CP-even observables}
Now we want to explore possible correlations between CP-even and odd observables, in particular, the correlation between Higgs observables and EDM experiments. To enhance the Higgs diphoton coupling, electroweak symmetry breaking needs to contribute negatively to the charged matter mass. This can be realized without introducing any physical CP phase, for example, through a single scalar with a large negative quartic coupling $-\lambda_S|S|^2|H|^2$ and $\lambda_S>0$ as already discussed in Sec.~\ref{sec:allscalar}. More generally, however, this is realized with presence of new CP phases, for example, in models with vector-like matter fields which obtain part of their masses from electroweak symmetry breaking. The general mass matrix, e.g., for fermions, is 
\beq\label{eq:MM}\mathcal{L}_M=-\left(\psi^{+Q}\;\chi^{+Q}\right)\left(\bear{cc}m_\psi&\frac{yv}{\sqrt{2}}\\\frac{y^cv}{\sqrt{2}}&m_\chi\ear\right)\left(\bear{c}\psi^{-Q}\\\chi^{-Q}\ear\right)+cc,\eeq
with the Higgs VEV given by $\langle H\rangle=v/\sqrt{2}=174$~GeV and $\psi, \chi$ are Weyl fermions. There is one physical phase, $\phi=\arg\left(m^*_\psi m^*_\chi yy^c\right)$, that cannot be rotated away by field redefinitions. In terms of operators, the diagrams generating CP-even operators, ${\cal O}_W, {\cal O}_{hW}, {\cal O}_{hB}$ lead to ${\cal O}_{h\tilde{W}}, {\cal O}_{h\tilde{B}}$ with insertion of the physical phase. Notice that the $WW\tilde{W}$ operator is not generated at one-loop. The reason is that the $W$'s and the $Z$ only couple to fermions of the same chirality. Without Higgs insertions, as each mass insertion flips chirality, the diagram is always proportional to even powers of $|m_\psi|^2$ or $|m_\chi|^2$, which are always real.  The $WW\tilde{W}$ operator could be generated at the two-loop order or, similar to the Weinberg operator $GG\tilde{G}$, $WW\tilde{W}$ receives a finite threshold correction from a heavy $SU(2)_W$ charged particle with a non-zero EDM $d_e$ and mass $m$,
\beq
a_{\tilde{W}}=-\frac{g^2}{96\pi^2}\frac{d_e}{s_WT_3m},
\eeq
where $g$ is the $SU(2)_W$ coupling. The constraint on $a_{\tilde{W}}$ from EDM translates into $\left|\frac{d_e}{T_3m}\right|<\frac{5 \times 10^{-17}e\cdot\,{\rm cm}}{1\,{\rm TeV}}$. 

\subsection{EDM}
As shown in Table~\ref{tab:constraints}, EDMs are more powerful CP-odd observables compared to the CP-odd TGC measurements. The current bound on the electron EDM is $d_e/e < 1.05 \times 10^{-27}$ cm, which will be improved by an order of magnitude in the near future~\cite{Vutha:2009ux}. If a non-zero electron EDM is observed then, there are three possibilities:
\begin{enumerate}
\item One-loop EDM if the charged matter has lepton quantum numbers with small CP phases ($\lsim 10^{-2}$); 
\item Higher-order contributions that will also affect the Higgs decays: in terms of high-dimensional operators, ${\cal O}_{h\tilde{W}}, {\cal O}_{h\tilde{B}}$ and their CP-even counterparts are generated. This could originate from vector-like charged matter with chiral mass terms;
\item Higher-order contributions that do not affect the Higgs decays, e.g., only $WW\tilde{W}$ is generated, perhaps from a heavy weakly-charged particle with a non-zero EDM as discussed in the previous subsection.
\end{enumerate}

In this section, we focus on the second possibility. We evaluate the EDM bounds and reach using the two-loop EDM formula in~\cite{Giudice:2005rz, Li:2008kz} and the modification of Higgs decaying to diphotons 
in two concrete models as discussed in~\cite{ArkaniHamed:2012kq}. The two models are 
\paragraph{Vector doublets + singlets (``vector-like lepton"): $\psi,\psi^c\sim(1,2)_{\pm\frac{1}{2}},\;\;\chi,\chi^c\sim(1,1)_{\mp1}.$} 
The Lagrangian leading to (\ref{eq:MM}) is
\ba\label{eq:doubL}-\mathcal{L}=m_\psi\psi\psi^c+m_\chi\chi\chi^c+yH\psi\chi+y^cH^\dag\psi^c\chi^c+cc.\ea
\paragraph{Vector doublets + triplet (``wino-higgsino"):  $\psi,\psi^c\sim(1,2)_{\pm\frac{1}{2}},\;\;\chi\sim(1,3)_{0}.$}
We identify $\chi$ and $\chi^c$; the Lagrangian leading to (\ref{eq:MM}) is
\ba-\mathcal{L}=m_\psi\psi\psi^c+\frac{1}{2}m_\chi\chi\chi+\sqrt{2}yH\psi\chi+\sqrt{2}y^cH^\dag\psi^c\chi+cc.\ea
In both cases, we define $\phi=$arg$(yy^cm_\psi^* m_\chi^*)\in (0, \pi/2)$ and $N$ the number of species for these fermions. 

\begin{figure}[!h]\begin{center}
\includegraphics[width=0.4\textwidth]{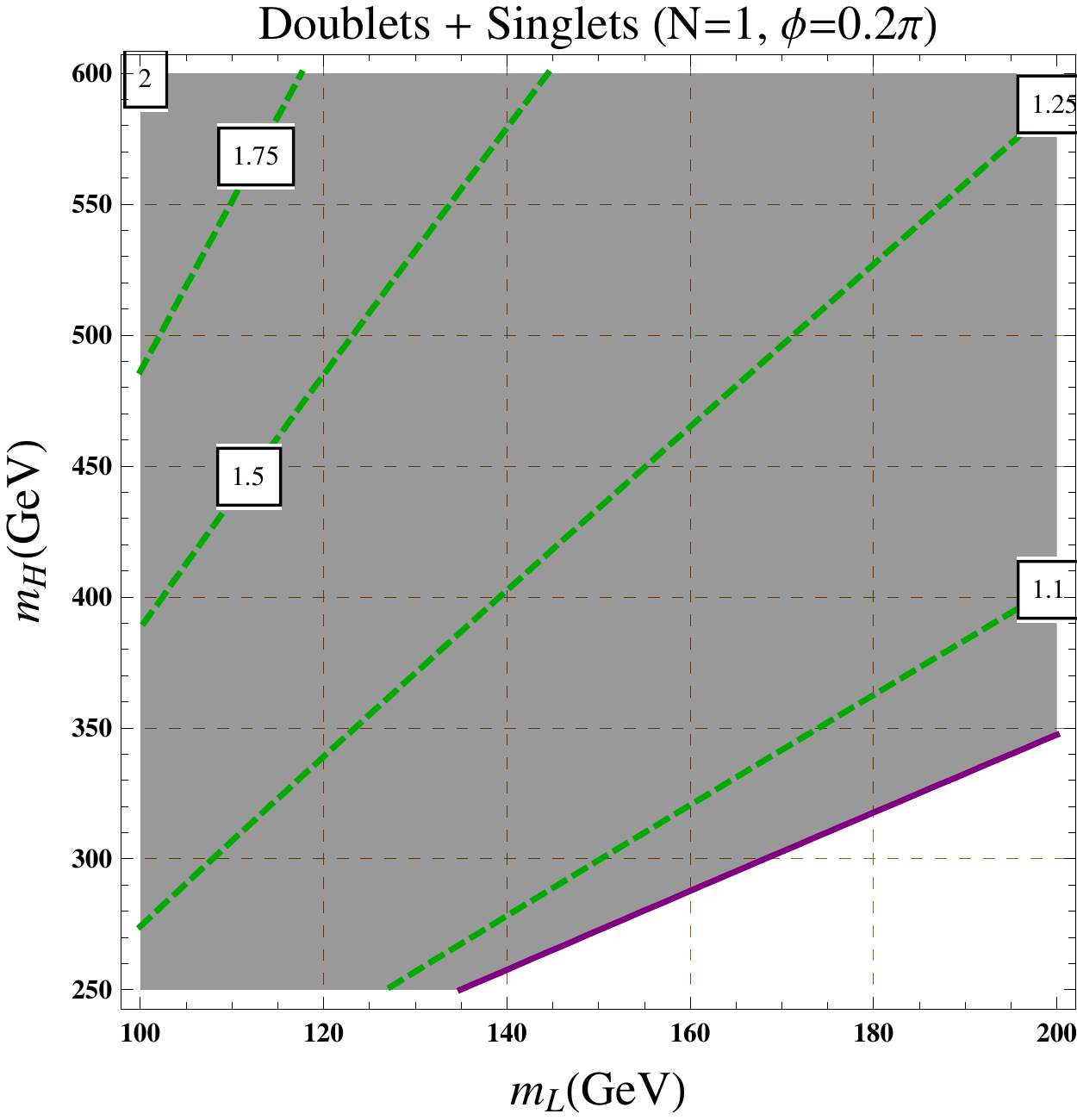}  \quad \includegraphics[width=0.4\textwidth]{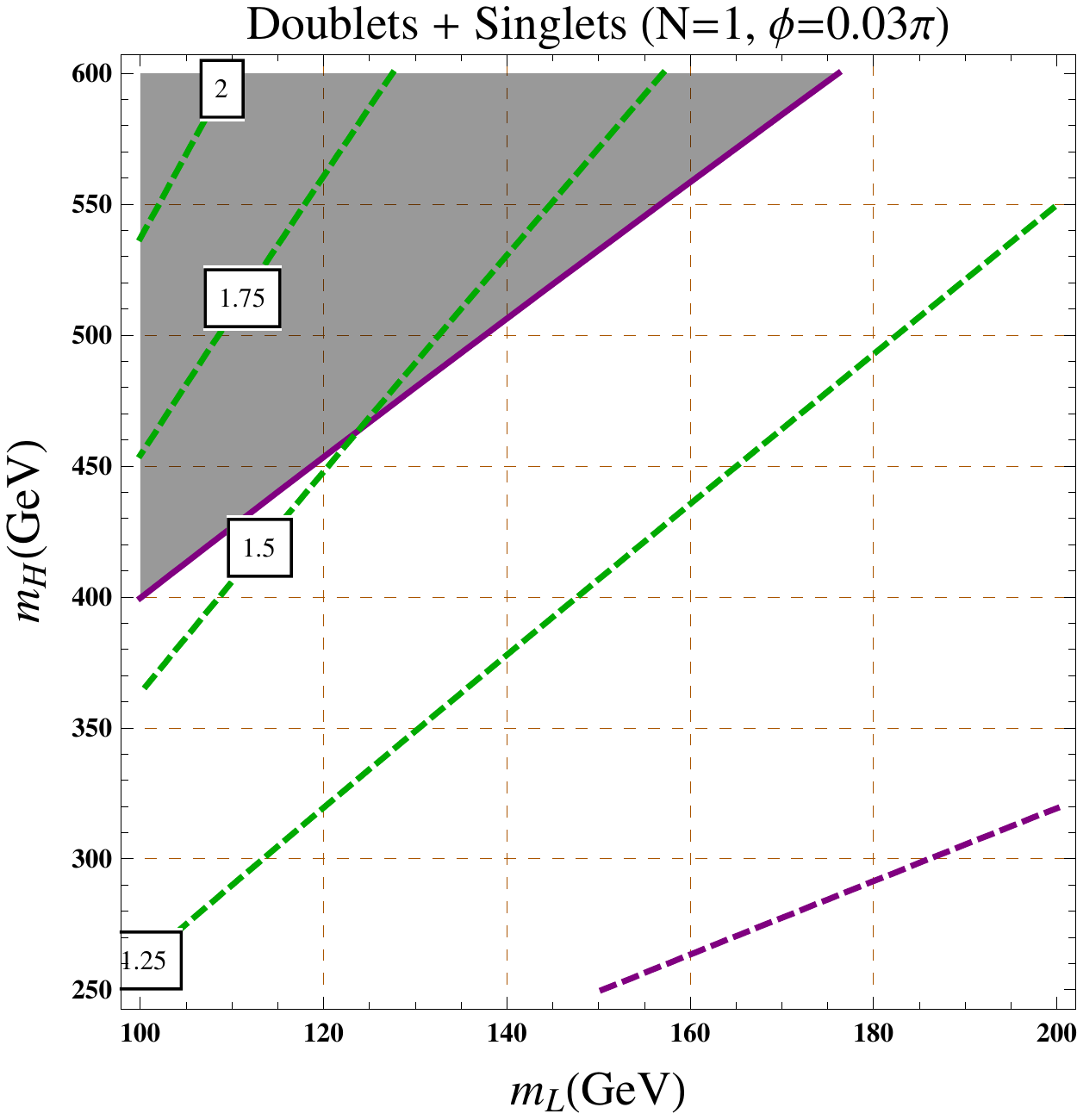}  \\
\includegraphics[width=0.4\textwidth]{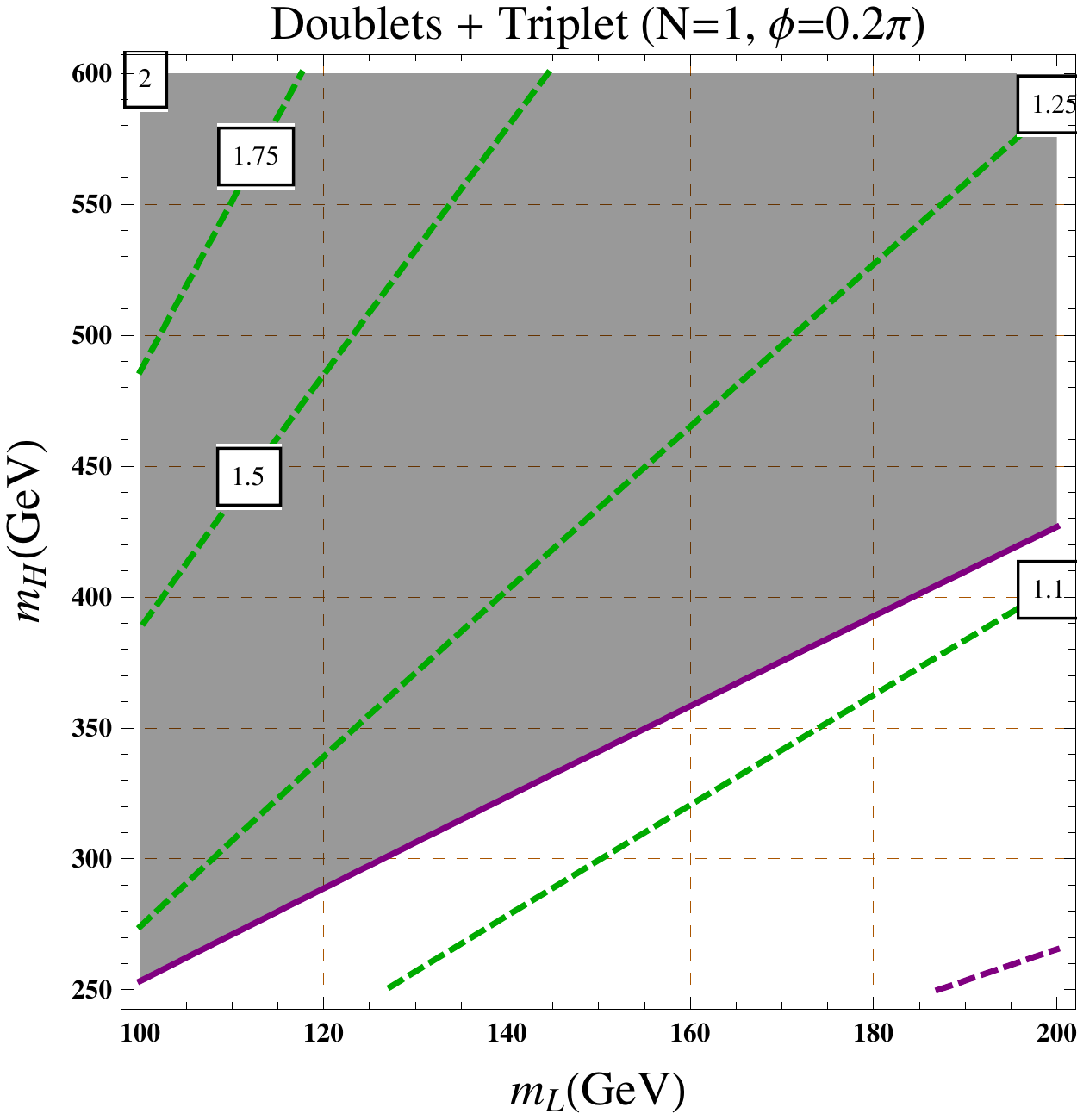}  \quad \includegraphics[width=0.4\textwidth]{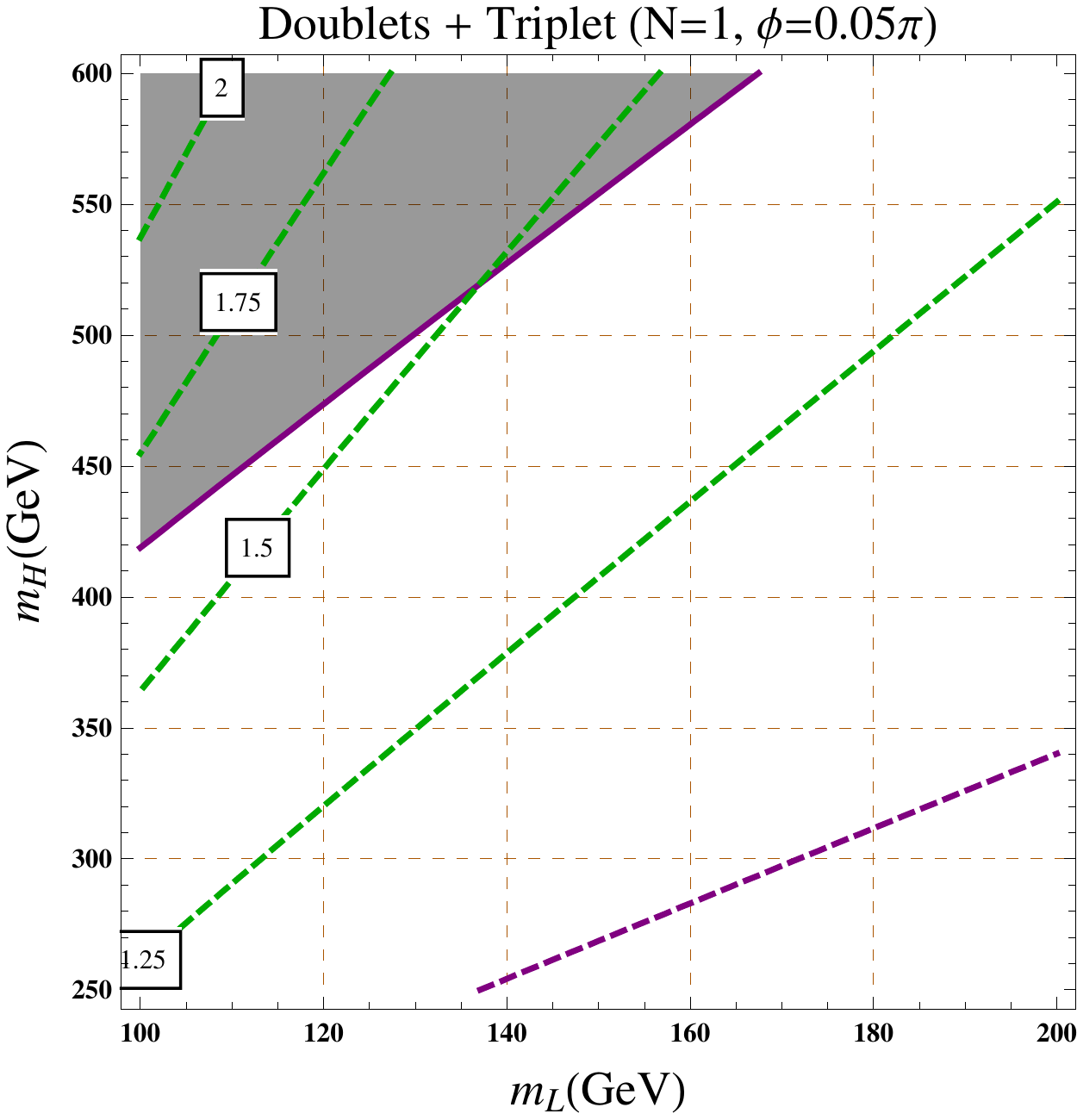}
\end{center}
\caption{Upper: ``vector-like lepton" model; Lower: ``wino-Higgsino" model. $N=1, m_\psi=m_\chi, y=y^c$ in all these plots. $\phi=$arg$(yy^cm_\psi^* m_\chi^*)$. The horizontal and vertical axes correspond to the light and heavy mass eigenvalues. The solid purple line is the current EDM constraint $d_e/e = 1.05 \times 10^{-27}$ cm with the grey region excluded; the dashed purple line is the projected constraint $d_e/e = 10^{-28}$ cm. The green lines denote the diphoton enhancement $\mu_{\gamma\gamma}$.}
\label{fig:bounds1}
\end{figure}%

\begin{figure}[!h]\begin{center}
\includegraphics[width=0.4\textwidth]{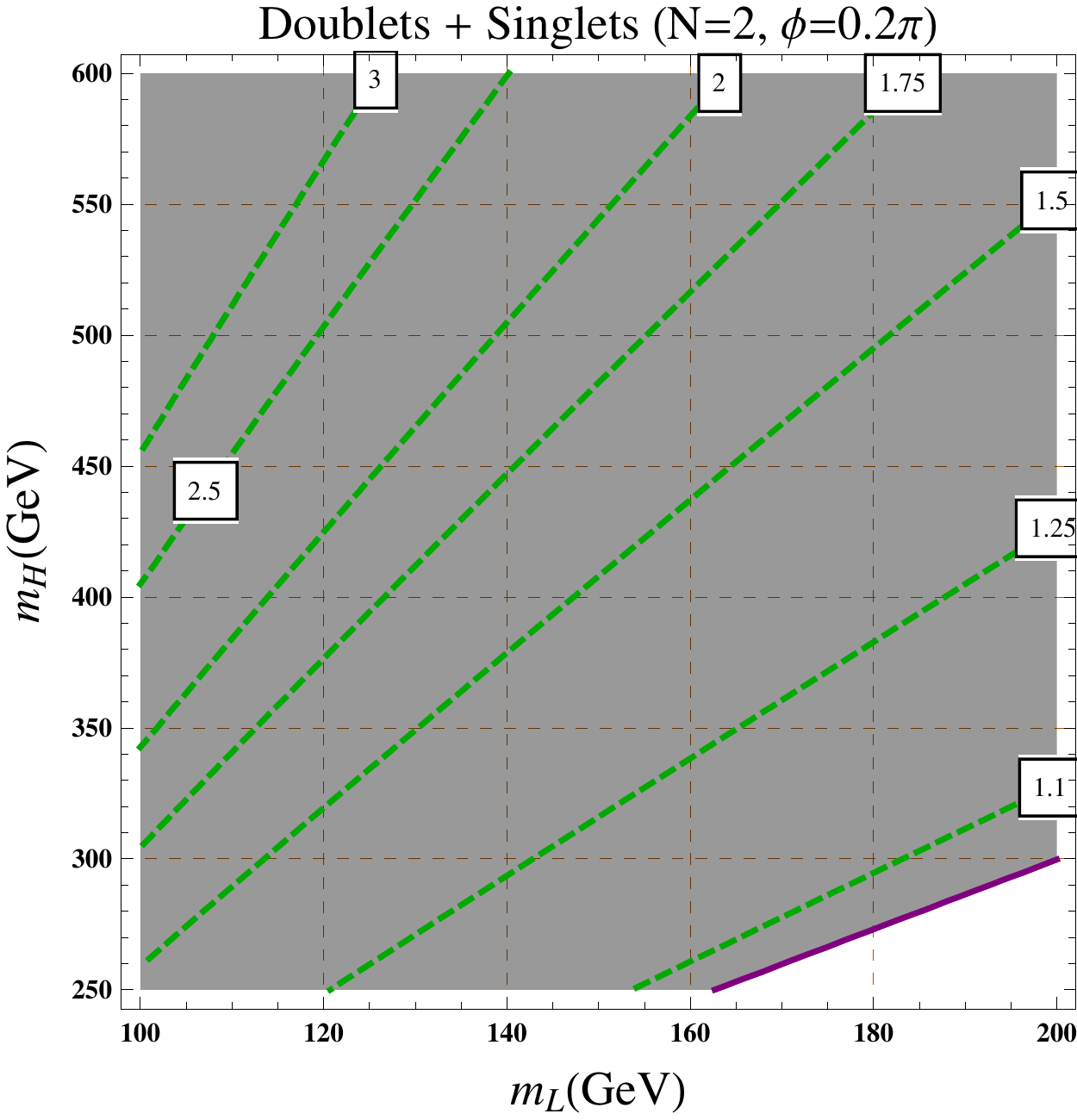}  \quad \includegraphics[width=0.4\textwidth]{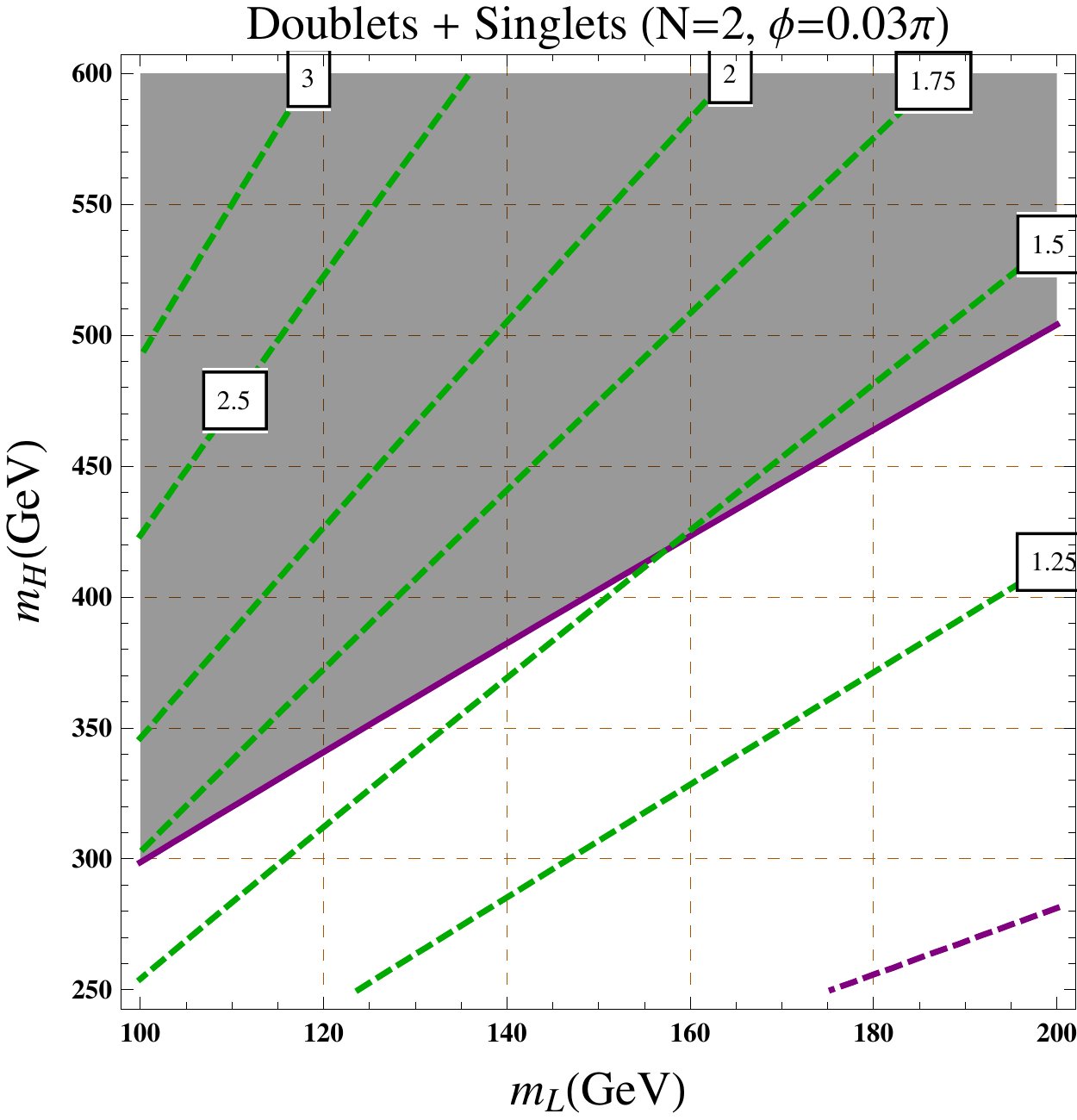}  \\
\includegraphics[width=0.4\textwidth]{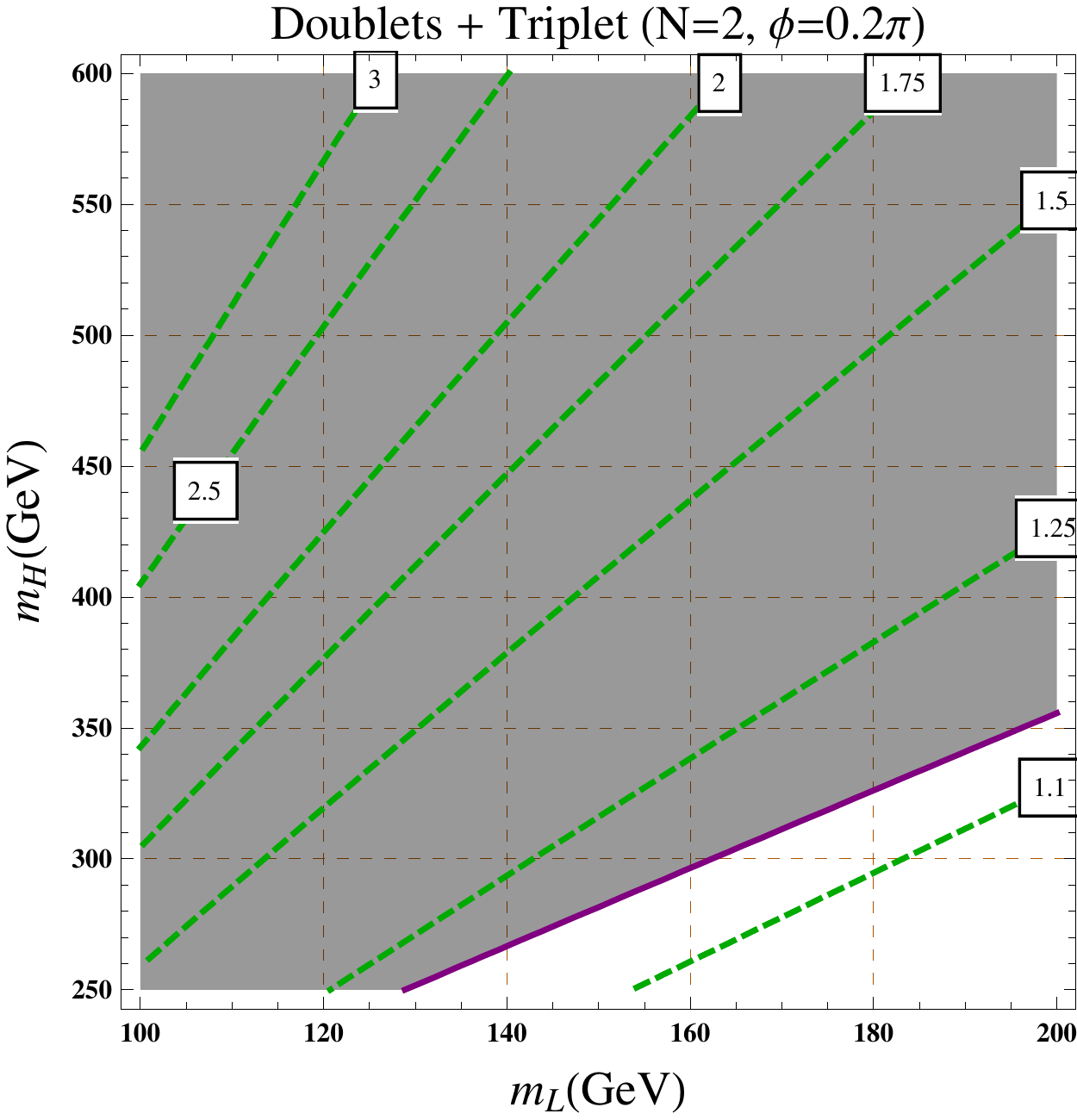}  \quad \includegraphics[width=0.4\textwidth]{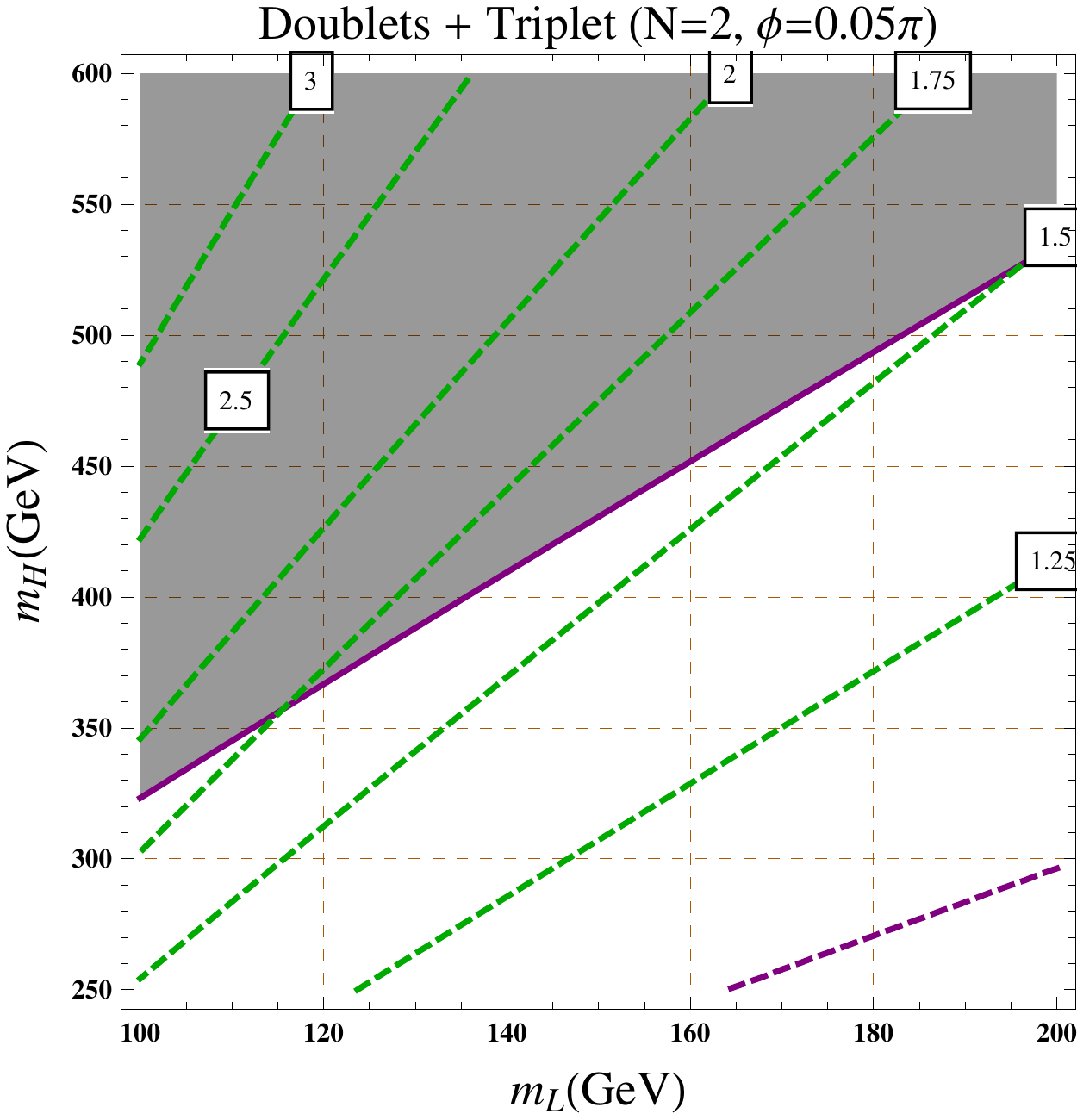}
\end{center}
\caption{Upper: ``vector-like lepton" model; Lower: ``wino-Higgsino" model. $N=2, m_\psi=m_\chi, y=y^c$ in all these plots. The horizontal and vertical axes correspond to the light and heavy mass eigenvalues. The solid purple line is the current EDM constraint $d_e/e = 1.05 \times 10^{-27}$ cm with the grey region excluded; the dashed purple line is the projected constraint $d_e/e = 10^{-28}$ cm. The green lines denote the diphoton enhancement $\mu_{\gamma\gamma}$.}
\label{fig:bounds2}
\end{figure}%

The results are shown in Figs.~\ref{fig:bounds1} and \ref{fig:bounds2}. To evade the stringent EDM bounds and have a diphoton enhancement $\mu_{\gamma\gamma} \geq 1.5$, the CP-violating phase has to be small: $\phi \leq 0.03 \pi$ for the ``vector-like lepton" model or $\phi \leq 0.05 \pi$ for the ``wino+Higgsino" model. Besides, for larger $\phi$, it is difficult to get a diphoton enhancement $\mu_{\gamma\gamma}$ as the enhancement mainly comes from the interference between the real part of the new fermion contribution and the SM contribution (for $N=1$).\footnote{For $N\geq 2$, the contribution from the imaginary part itself could be important as it scales as $N^2$ while the interference between the real parts scales as $N$.} For fixed light and heavy mass eigenvalues, the real part of the new fermion contribution, which scales with $|y y^c| = y y^c \cos \phi$, decreases as $\phi$ gets bigger. Thus if the diphoton enhancement is confirmed, it will pose a new interesting Higgs CP problem at the weak scale analogous to the SUSY CP problem: \textbf{Why do those new charged degrees of freedom that modify Higgs decaying to photons have small phases?}

Neutron EDMs lead to comparable constraints. The case where new charged particles are {\em scalars} (like staus) rather than fermions leads to a somewhat different story. Even if there are nontrivial phases in their mass matrix (e.g., in the $A$-terms in the case of staus), integrating out such particles does not generate the operator $hF{\tilde F}$. Although this can be straightforwardly seen from the loop calculation, it can also be understood as a consequence of the fact that the $\arg\det{\cal M}$ coupling arises from an anomalous rotation of fermion fields, whereas scalars have no anomalies. However, if there is a pseudoscalar particle in the spectrum that can run in the two-loop EDM diagram in place of the Higgs, or if CP-violation leads the Higgs to have a small pseudoscalar-like coupling to the electron (e.g. by mixing with a pseudoscalar), there will still be a two-loop EDM~\cite{Chang:1998uc}. Thus, in the case of charged scalars, the Higgs CP problem would be less robust: if all pseudoscalars are heavy, the EDMs can be rather small. (There are other difficulties for such an interpretation of an increased $h \to \gamma\gamma$ rate, as new charged scalars typically have vacuum stability problems~\cite{Reece:2012gi,Kitahara:2012pb}, although there is still viable parameter space for quite light scalars~\cite{Carena:2012mw}.)

\section{Conclusions}
\label{sec:conclusions}
Charged weak-scale matter is the key ingredient in most explanations of  $h\to\gamma\gamma$ enhancement and a monochromatic photon line at around 130 GeV in the Fermi data. It is tempting to have a unified explanation for both of them in which some charged matter couples to both Higgs and DM. We find that a fine-tuning at 10\% level or worse is inevitable in these models. The large couplings required by the $h\to\gamma\gamma$ enhancement and the photon flux responsible for the line would lead to low-scale Landau poles. Even worse, they could induce a considerable DM-Higgs coupling radiatively even if this coupling vanishes at tree level. To evade the direct detection and photon continuum constraints, one needs to invoke a 10\% level tuning between the tree-level and radiative DM-Higgs coupling. In certain models such as the resonant annihilation models, the large couplings and Landau poles could be avoided by tuning the mass of the intermediate boson to be close to twice the DM mass for no deep reason. For example, with a 10\% coincidence in mass, the required couplings are approximately 1. Constraints on DM-Higgs couplings from direct detection could be avoided in models with DM as Majorana fermions and small CP violating phases. In light of these and other difficulties, like the vacuum stability problems that plague $h\to\gamma\gamma$ models, it seems likely that one or both signals will fade away; the alternative is not only that light charged particles exist, but that a spectacular array of interesting new physics will be found at nearby energies. A more optimistic way of stating this result independent of observed anomalies is that the current experimental sensitivity to $h \to \gamma\gamma$, gamma-ray lines from dark matter annihilation, and dark matter couplings to the Higgs induced at one loop by charged weak-scale particles are all comparable, so that if new particles are lurking just below the current limits, we can hope that a spectacular set of correlated signals will emerge.

A new CP violating phase could also be present in the couplings of new charged weak-scale matter. Even if the new matter does not couple or mix with the SM fermions at tree-level, they could generate two-loop EDMs through Barr-Zee type diagrams. In particular, in models with new vector-like fermions to enhance the diphoton rate to be more than 1.5 times the SM rate, an order-one new CP phase leads to too large an EDM that is ruled out already unless a cancelation with other contributions, corresponding to a few percent tuning, is present. This will impose an interesting Higgs CP problem, that is, if the diphoton enhancement is true and caused by charged fermions, the new CP phase has to be about 0.1 or less. An ongoing electron EDM experiment will improve the electron EDM bound by about one order of magnitude in the near future, potentially worsening the Higgs CP problem. In models with charged scalars to enhance Higgs diphoton rate, the two-loop EDM could be negligible if there is no light pseudoscalar mixing with the Higgs. Thus, in the case of charged scalars, the Higgs CP problem would be less robust. However, vacuum instability problems constrain the charged scalar models severely. 

It is remarkable that experiments underground (direct detection), in atomic physics laboratories (EDM measurements), in colliders, and in space (Fermi-LAT) are setting comparable bounds on new charged matter. This offers the prospect of a stunning coincidence of signals, if such matter exists.

\acknowledgments{JF, MR are supported in part by the Fundamental Laws Initiative of the Harvard Center for the Fundamental Laws of Nature. JF acknowledges the hospitality of the Aspen Center for Physics, which is supported by the National Science Foundation Grant No. PHY-1066293. MR thanks the Galileo Galilei Institute for Theoretical Physics in Florence, Italy, for its hospitality while a portion of this work was completed.}

\begin{appendix}

\section{Fermionic DM through a box diagram}
\label{app:boxdiagram}

In this appendix we will summarize some results on one further model, to show that the qualitative conclusions reached in Section~\ref{sec:allscalar} are not strongly dependent on the initial assumption that all particles involved are scalars. We assume DM is a fermion $\xi, \xi^c$ annihilating through loops of charged fermions $\psi, \psi^c$ and a charged scalar $\phi$:
\beq
m_\phi^2 \left|\phi\right|^2 + \left(g_\xi \phi \psi^c \xi + g_\xi^c \phi^\dagger \xi^c \psi + m_\xi \xi^c \xi + m_\chi \chi^c \chi + m_\psi \psi^c \psi + y H\psi \chi^c+ y^c H^\dagger \psi^c \chi + {\rm h.c.}\right).
\eeq
The fermions $\psi,\chi$ are, as before, a singlet and doublet, while $\phi$ is a charge 1 scalar. In this case the RGEs are:
\beq
16\pi^2 \beta(g_\xi) = 2 g_\xi^3 + g_\xi^{c2} g_\xi + y^{c2} g_\xi, \\
16\pi^2 \beta(y) = \frac{5}{2} y^3 + y^{c2} y + g^{c2}_\xi y,
\eeq
together with the analogous equations where couplings with and without superscript $c$ are interchanged. We fix the masses $m_\psi$, $m_\chi$ and the Yukawas $y, y^c$ as in Section~\ref{sec:resonantannihilation} to give a 50\% enhancement of the $h \to \gamma\gamma$ rate. In this case we don't need to include a CP-violating phase, so we take $m_\chi = m_\psi = 335$ GeV, $y = y^c = 1.12$, which produces charged fermion mass eigenstates at $m_{1,2} = 140,\, 530$ GeV. The mass matrix for the fermions is diagonalized by two matrices $U$ and $V$, i.e. $U^* {\cal M} V^\dagger$ is the diagonal matrix with eigenvalues $m_1$ and $m_2$ (as in, e.g., Ref.~\cite{ArkaniHamed:2004yi}).

We fix $m_\xi = 130$ GeV. For simplicity we will fix $g_\xi^c = g_\xi^\dagger$. Then the annihilation rate through a loop of charged fermions and scalars is given by~\cite{Bergstrom:1997fh,Bern:1997ng,Tulin:2012uq}:
\beq
\sigma(\xi\xi^c \to \gamma\gamma)v &= & \frac{\alpha^2 m_\xi^2}{64\pi^3 m_\phi^4} \left|\sum_i g_\xi^2 U_{\psi^c i}V_{\psi i} \frac{m_\phi^2}{m_i^2} \left(\frac{m_\xi^2 + m_\xi m_i}{m_\phi^2 + m_\xi^2 - m_i^2} I_1\left(\frac{m_\xi^2}{m_\phi^2},\frac{m_i^2}{m_\phi^2}\right) + \frac{m_\phi^2}{m_\phi^2 - m_i^2} I_2\left(\frac{m_\xi^2}{m_\phi^2},\frac{m_i^2}{m_\phi^2}\right) \right.\right. \nonumber \\
& & \quad\quad \left.\left.+ \left(\frac{2 m_i^2 + 2 m_i m_\xi}{m_\phi^2 + m_\xi^2 - m_i^2} - \frac{m_i^2}{m_\phi^2 - m_i^2}\right) I_3\left(\frac{m_\xi^2}{m_\phi^2},\frac{m_i^2}{m_\phi^2}\right)\right)\right|^2,
\eeq
with the loop functions defined in Ref.~\cite{Bergstrom:1997fh}. Then we can achieve a dark matter annihilation to diphoton rate of $10^{-27}~{\rm cm}^3/{\rm s}$ with the choice: $m_\phi = 150$ GeV, $g_\xi = 3.6$. As in the case of an all-scalar loop, this is a rather large coupling; in fact, with our definition of normalized coupling from Sec.~\ref{subsec:scalarrges}, solving the RGE shows that $g_\xi$ and $g^c_\xi$ reach a normalized value of 1 already at a renormalization scale of 235 GeV. They reach a value of 4$\pi$ well below a TeV. Hence, in the case of a box diagram, one really should imagine that the theory is one of composite dark matter coupling to composite charged fields nearby in mass, much as in the all-scalar case we considered above.

In this case there is again a loop-induced dark matter coupling to the Higgs. We have not fully calculated the diagram, but schematically one expects a coupling of the form
\beq
\frac{g_\xi g_\xi^c y y^c m_\chi}{16\pi^2 \Lambda^2} H^\dagger H \xi \xi^c,
\eeq
with $\Lambda$ of order the masses of particles running in the loop. Taking $\Lambda \sim 150$ GeV and plugging in our choices, we find a coupling $\sim 0.4 h \xi \xi^c$, in conflict with direct detection bounds unless the numerical coefficient proves to be quite small. We expect that a more careful calculation would continue to support the same qualitative story as in Section~\ref{sec:allscalar}: without resonant enhancement, the couplings involved in fitting the line rate and a large $h \to \gamma\gamma$ excess are so large as to pose a strong tension with the absence of large direct dark matter couplings to the Higgs implied by XENON.

\section{Counting CP-odd operators}
\label{app:CPoddops}
The literature can be slightly confusing on the enumeration of the set of dimension-six CP-odd operators constructed from electroweak gauge bosons and the Higgs. In particular, one can find two more CP-odd dimension six operators besides the ones we listed in Sec.~\ref{sec:ops}: $(D_\mu H)^\dagger \sigma^a D_\nu H\tilde{W}_{\mu\nu}^a$ and $(D_\mu H)^\dagger  D_\nu H\tilde{B}_{\mu\nu}$~\cite{Giudice:2005rz, Jung:2008it}. We will show that these two operators are related by equations of motion to the ones we listed. We will take $(D_\mu H)^\dagger \sigma^a D_\nu H\tilde{W}_{\mu\nu}^a$ for example. 
\beq
&&(D_\mu H)^\dagger \sigma^a D_\nu H\tilde{W}^{\mu\nu;a}=(\partial_\mu H^\dagger +i \frac{g}{2}H^\dagger W_\mu^b\sigma^b+i\frac{g^\prime}{2}H^\dagger B_\mu Y)\sigma^a D_\nu H\tilde{W}^{\mu\nu;a} \nonumber\\
&=&- H^\dagger \sigma^a D_\nu H  \partial_\mu \tilde{W}^{\mu\nu;a} - H^\dagger \sigma^a (\partial_\mu D_\nu H)\tilde{W}^{\mu\nu;a} +\left(i \frac{g}{2}H^\dagger W_\mu^b\sigma^b+i\frac{g^\prime}{2}H^\dagger B_\mu Y\right)\sigma^a D_\nu H\tilde{W}^{\mu\nu;a} \nonumber\\
&=&-H^\dagger \sigma^a D_\nu H  \partial_\mu \tilde{W}^{\mu\nu;a} -H^\dagger \sigma^a \left(\left(\partial_\mu  +i\frac{g}{2}W_\mu^b\sigma^b+i\frac{g^\prime}{2}B_\mu Y\right)D_\nu H \right)\tilde{W}^{\mu\nu;a}+gH^\dagger \epsilon_{abc}W_\mu^b\sigma^cD_\nu H\tilde{W}^{\mu\nu;a} \nonumber \\
&=&-H^\dagger \sigma^a D_\nu H  (-g\epsilon_{abc}W_\mu^b\tilde{W}^{\mu\nu;c}) -H^\dagger \sigma^a (D_\mu D_\nu H)\tilde{W}^{\mu\nu;a}+gH^\dagger\epsilon_{abc}W_\mu^b\sigma^cD_\nu H\tilde{W}^{\mu\nu;a}  \nonumber \\
&=&-\frac{1}{2}H^\dagger \sigma^a [D_\mu, D_\nu]H\tilde{W}^{\mu\nu;a}\nonumber \\
&=&\frac{i}{4} H^\dagger \sigma^a (g W_{\mu\nu}^b\sigma^b+g^\prime B_{\mu\nu})H \tilde{W}^{\mu\nu;a}=\frac{i}{4} g H^\dagger H W_{\mu\nu}\tilde{W}^{\mu\nu}+\frac{i}{4} g^\prime H^\dagger \sigma^a H  B_{\mu\nu}\tilde{W}^{\mu\nu;a},
\eeq
where in the second line we integrated by parts; in the third line we used $[\sigma^a, \sigma^b]=2i\epsilon_{abc}\sigma^c$; in the third line we used Bianchi identity $\epsilon^{\mu\nu\rho\sigma}(D_\mu W_{\rho\sigma})^a=0$; in the last line we used $\sigma^a\sigma^b=\delta^{ab}I+i \epsilon^{abc}\sigma^c$.
\end{appendix}

\bibliography{ref}
\bibliographystyle{jhep}
\end{document}